\documentclass[a4paper,11pt]{article}
\usepackage{xcolor}
\usepackage[normalem]{ulem}
\usepackage{textcomp}
\pdfoutput=1 

\usepackage{jcappub} 

\usepackage[T1]{fontenc} 

\usepackage{caption}
\usepackage{subcaption}

\definecolor{ao(english)}{rgb}{0.0, 0.5, 0.0}

\title{Constraints on the growth of structure around cosmic voids in eBOSS DR14}

\author[a,1]{Adam J. Hawken\note{Corresponding author.},}
\author[a]{Marie Aubert,}
\author[b]{Alice Pisani,}
\author[a]{Marie-Claude Cousinou,}
\author[a]{Stephanie Escoffier,}
\author[c]{Seshadri Nadathur,}
\author[d]{Graziano Rossi,}
\author[e,f]{Donald P. Schneider}

\affiliation[a]{Aix Marseille Univ, CNRS/IN2P3, CPPM, Marseille, France}
\affiliation[b]{Department of Astrophysical Sciences, Princeton University, Princeton, New Jersey, USA}
\affiliation[c]{Institute of Cosmology and Gravitation, University of Portsmouth, Burnaby Road, Portsmouth PO1 3FX, UK}
\affiliation[d]{Department of Physics and Astronomy, Sejong University, Seoul, 143-747, Korea}
\affiliation[e]{Department of Astronomy and Astrophysics, The Pennsylvania State University,
   University Park, PA 16802, USA}
\affiliation[f]{Institute for Gravitation and the Cosmos, The Pennsylvania State University,
   University Park, PA 16802, USA}

\emailAdd{hawken@cppm.in2p3.fr}

\abstract{We present catalogues of cosmic voids identified in the distribution of Luminous Red Galaxies (LRGs) and Quasi Stellar Objects (QSOs) in the fourteenth data release (DR14) of the extended Baryon Oscillation Spectroscopic Survey (eBOSS). We perform a multivariate analysis to assess the level of contamination in these catalogues by spurious Poisson underdensities. We find that the LRG void catalogue is largely free from contamination but that the QSO catalogue may be heavily contaminated. We analyse the multipoles of the void-galaxy cross-correlation function in these catalogues to obtain constraints on the growth rate of structure around voids. We find a value of $\beta(z=0.703)=0.58^{+0.33}_{-0.28}$ for the LRG voids and $\beta(z=1.53)=0.15^{+0.13}_{-0.12}$ for the QSO voids. }

\begin{document}

\maketitle

\flushbottom

\section{Introduction}
\label{sec:intro}

Cosmic voids are large underdense regions of the Universe, ranging in size from tens to hundreds of $h^{-1}$Mpc. Together with clusters, walls, and filaments they make up the cosmic web. Voids have been known about since the pioneering galaxy redshift surveys of the 1970s \citep[e.g.][]{GregoryThompson1978}. 
It has long been speculated that the abundance and dynamical properties of these structures carry useful cosmological information
\citep[e.g.][]{Sheth2004}.
It is only relatively recently though, with the advent of large volume galaxy redshift surveys and the subsequent increase in statistics, that their potential as a powerful cosmological probe has become apparent \citep[e.g.][and references therein]{LavauxWandelt2012, Pisani2019}.

The Sloan Digital Sky Survey \citep[SDSS]{York2000, Eisenstein2011, Blanton2017} has proven to be a valuable resource  
for the study of cosmic voids. Cosmic voids in this survey have been utilised for many cosmological applications and have yielded several key results. 
Among them, measurements of the velocity and density profiles of cosmic voids \cite{Paz2013, Hamaus2016}; the impact of survey masks on void properties \citep{Sutter2014}; the void auto-correlation function and clustering bias \citep{Clampitt2016}; the depth, abundance, 
and general properties of voids \citep{Sutter2012, Nadathur2014,  Mao2017cats, Nadathur2016, Nadathur2019}; the Alcock-Paczynski test \citep{Sutter2012AP, Mao2017, Hamaus2016}.
The properties of voids in the CMASS population in particular, luminous galaxies in the redshift range $0.4 \leq z \leq 0.7$ targeted by the Baryon Oscillation Spectroscopic Survey (BOSS \citep{Dawson2013}, part of SDSS-III \citep{Eisenstein2011}), have been well studied. A number of studies have been made of voids in the CMASS DR12 data set \citep{Cai2017, Mao2017, Cousinou2019, Hamaus2017Multipoles, Nadathur2019, Achitouv2019, Nadathur2016ISW, Kovacs2019, Alonso2018}. Several of these studies also used the LOWZ data set, i.e. BOSS targets at low redshift (z <~ 0.4).


Since the days of Hubble
, and the establishment of the redshift-distance relationship, the recessional velocities of galaxies have been used as a proxy for their distance. Galaxies are, however, subject to motions apart from the Hubble flow.  
These peculiar motions add an additional Doppler component to the redshift and thus distort the apparent positions of galaxies in redshift space. These Redshift Space Distortions (RSD) introduce anisotropies to the observed clustering pattern of galaxies. 
At the linear level, 
galaxy peculiar motions are caused by galaxies falling onto overdense structures. The strength of RSD is therefore closely related to the growth rate of cosmic structure; thus isolating this signal can provide key insights into the nature of gravity.

The strength of linear RSD is related to the growth rate of cosmic structure and can be quantified using the parameterisation  $\beta(z)=\Omega_m(z)^\gamma/b$, where $\Omega_m(z)$ is the matter density of the Universe at redshift, $z$; $b$ is the galaxy bias; and the exponent $\gamma$ depends on the theory of gravity (in standard General Relativity, GR, $\gamma\approx0.55$ \citep{Lahav1991}), and likewise on the dark energy equation of state \citep{Linder2003}. Thus measuring $\beta$ is one of the ways in which we can distinguish between the $\Lambda$CDM paradigm and alternative cosmic scenarios. The phenomenon of RSD has been known about for more than $30$ years \citep{Kaiser1987, Hamilton1992, Peacock2001}
, but their usefulness for probing gravity was only realised a decade ago \citep{Guzzo2008}.

Conventional measurements of the growth rate are derived from observations of either the galaxy auto-correlation function \citep[e.g.][]{Zarrouk2018}, or the galaxy power spectrum \citep[e.g.][]{Zhao2019}. In principle these approaches are equivalent, though in practice they are subject to quite different systematics. Chief among the difficulties with these methods is the non-linear nature of small-scale gravitational interactions \citep{Peebles1980, PeacockDodds1994, Fisher1995}, which become important when the precision of clustering measurements is better than 10\% \citep{Okumura2011, Bianchi2012}. 


The velocity field in and around cosmic voids is dominated by coherent bulk flows \citep{Dubinski1993, Aragon-Calvo2013}. The density of material close to the edges of cosmic voids is the same order of magnitude as the mean cosmic density. Therefore the relationship between the matter density field and the velocity field, in these regions, is well described by linear theory \citep{Hamaus2013Universal}. Furthermore, many modified gravity models predict that gravitational dynamics might deviate from standard GR more strongly inside voids than in denser environments \citep[e.g.][]{HuSawicki2007}. 
Thus accurate measurements of the growth rate of structure around cosmic voids are complementary to standard techniques and allow us to place constraints on gross deviations from standard GR \citep{Hamaus2016, Hawken2017, Hamaus2017Multipoles, Achitouv6dF, Achitouv2019, Nadathur2019}. 

The first model used to describe RSD around voids was the Gaussian
streaming model \citep{Fisher1995}. It posits that the flow of matter from void interiors onto surrounding structures is coherent and has a Gaussian velocity dispersion. Ref. \citep{Paz2013}
demonstrated that this model can be used to describe anisotropies observed in the void-galaxy cross-correlation function in SDSS.
Ref. \citep{Hamaus2015}
showed using voids in N-body simulations that, in combination with a model for the void density profile and accounting for the Alcock--Paczynski effect, the Gaussian streaming model could be used to make a measurement of the growth rate of structure; demonstrating its applicability to SDSS data \citep{Hamaus2016}. Ref. \citep{Achitouv6dF} applied a similar model to voids in the 6dF survey \citep{6DF}. One of the drawbacks of the Gaussian streaming model is that it requires prior knowledge of the distribution of matter around voids (the void density profile, also dependant on cosmology), which in the aforementioned studies was provided by a model that was marginalised over. Ref. \citep{Hawken2017} applied the Gaussian streaming model to voids identified in VIPERS \citep{Guzzo2014}. By deprojecting the cross-correlation to estimate the void density profile, they successfully obtained a measurement of the growth rate of structure. At the time of writing, this result is the highest redshift measurement of the growth rate using voids. 

It has been suggested that the Gaussian streaming model may have a number of potential problems. Ref. \cite{Hamaus2015} found that the Gaussian streaming model can give large systematic errors on RSD and AP parameters. This behaviour was found to be particularly the case for smaller voids and denser tracer populations. For the case of a spherical void finder, Ref \cite{Cai2016} demonstrated that the Gaussian streaming model does not accurately describe RSD close to the centres of voids. Furthermore, the Gaussian streaming model contains a velocity dispersion parameter that must be marginalised over. Ref. \cite{NadathurPercival2017} have argued that these problems are due to the fact that the Gaussian streaming model is built by analogy with the galaxy auto-correlation function, and that the same rigorous derivation applied to the void-galaxy cross-correlation may not lead to a viable RSD model.

Recent developments in the methodology of studying RSD around voids may be more promising than the Gaussian streaming model. Measurements of the multipoles of the void-galaxy cross-correlation function in redshift space can be used to measure the growth rate of structure \citep{Cai2016}. This formulation has fewer parameters than the streaming model, also it does not require any prior knowledge of the void density profile. Ref.  \citep{Hamaus2017Multipoles} have applied the multipole technique to BOSS data, obtaining results consistent with previous measurements and competitive with standard RSD techniques. Refs. \cite{Nadathur2019} and \cite{Achitouv2019} also measured the multipoles in BOSS data but using a different technique. 
In this paper we will show how we used the multipoles of the cross-correlation between voids identified in the extended Baryon Oscillation Spectroscopic Survey \citep[eBOSS,][]{Dawson2016} large scale structure catalogues and tracers in these catalogues to obtain a measurement of the RSD parameter, $\beta$. 

To determine distances in cosmology it is first necessary to have a fiducial cosmological model that allows the observer to calculate quantities such as the angular diameter distance $d_A(z)$ and Hubble function $H(z)$. Inevitably the underlying model the cosmologist has is incorrect in some way. This results in geometric distortions known as the Alcock-Paczynski effect \citep{AlcockPaczynski1979} -- distances measured along the line-of-sight look different to those measured perpendicular to the line-of-sight. In the context of cosmic voids this can result in an apparent stretching or squashing along the line-of-sight, making otherwise spherical on average voids appear elliptical. The Alcock-Paczynski effect is degenerate with RSD. However, the void-galaxy cross-correlation is actually much better at measuring the Alcock-Paczynski effect than two-point galaxy clustering statistics \citep{Nadathur2018}. Therefore some authors argue that it is beneficial to study the two effects simultaneously. However, here we shall follow refs. \citep{Hawken2017, Hamaus2017Multipoles, Achitouv2019} and will not include Alcock-Paczynski parameters in our analysis, making the assumption that the Alcock-Paczynski effect is negligible; i.e. that our fiducial cosmology is reasonably close to the true cosmology and any discrepancy is too small to have a measurable impact on the calculation of cosmological distances. Inclusion of the Alcock-Paczynski effect is beyond the scope of this work. However, we plan to take this effect into account in future analyses of eBOSS data.

The paper proceeds as follows: In Section \ref{sec:data} we describe the selection of the samples in which we searched for voids and the production of the mock galaxy catalogues designed to emulate these samples. In Section \ref{sec:voidfinding} we give a brief overview of our void finding algorithm, and describe the summary statistics of our void catalogues. In Section \ref{sec:measurement} we describe how we measured the void-galaxy cross-correlation function, reviewing the linear RSD model for the multipoles of the void-galaxy cross-correlation function, and describing how we fitted this model to the observed cross-correlation to obtain an estimate of the growth rate of structure around cosmic voids. In Section \ref{sec:MVA} we describe how we applied Multivariate Analysis techniques to attempt to mitigate the systematic effects of spurious underdensities. In Section \ref{sec:results} we present our constraints on the growth rate of structure around cosmic voids. In Section \ref{sec:discussion} we discuss the significance of our results and how to take this work further.

\section{Data and mock catalogues}
\label{sec:data}
\subsection{Data}
eBOSS \citep{Dawson2016} is part of the Sloan Digital Sky Survey-IV (SDSS-IV) suite of astronomical surveys \citep{Blanton2017}. eBOSS uses the same optical spectrographs as the SDSS-III BOSS survey \citep{Dawson2013, Smee2013},
installed on the 2.5 meter Sloan Foundation Telescope \citep{Gunn2006} at the Apache Point Observatory in New Mexico. eBOSS targets four different types of tracer: Luminous Red Galaxies (LRGs), Emission Line Galaxies (ELGs), and 
two different quasar (QSO) populations, one as a direct clustering tracer and another for studies of the Lyman-$\alpha$ (Ly-$\alpha$) forest.  
Below we describe the two samples used in this analysis, namely, LRGs and QSO clustering targets.
The main characteristics of the eBOSS large scale structure catalogues are given in Table~\ref{table:numberofgal}.

\subsubsection{Luminous Red Galaxies}

Luminous Red Galaxies (LRGs) are the most massive type of galaxy. These passively evolving galaxies, as the name suggests, are also the most luminous and the reddest. LRGs are a well studied population \citep{PostmanLauer1995, York2000}. They are strongly clustered and tend to reside in massive dark matter haloes. Their bright intrinsic luminosity means that they can be observed over a wide redshift range. LRGs are also a remarkably uniform population in terms of their bias and their stellar composition. Therefore they are excellent tracers of the large scale structure.  The LRG catalogue used in this paper is the union of CMASS galaxies described in \citep{Reid2016}, above a redshift of $z>0.6$, and DR14 LRGs described in \citep{Prakash2016, Bautista2018}, covering the redshift range $0.6 \leq z \leq 1.0$.  

\subsubsection{Quasi Stellar Objects}

The vast majority of unresolved extragalactic objects that are bluer than main sequence stars are quasars \citep[Quasi Stellar Objects, QSOs][]{Sandage1965}. Quasars are extremely luminous active galaxies easily detectable at $z>1$. The quasar catalogue used in this paper \citep[DR14Q,][]{Paris2018} consists of around 150,000 quasars spanning the redshift range $0.8\leq z \leq2.2$. Conventional galaxy clustering measurements have been applied to this catalogue to measure the growth rate of structure \citep{Zarrouk2018, Gil-Marin2018}. Studies have been made on voids in the Ly-$\alpha$ forest \citep[e.g.][]{Krolewski2018}. eBOSS also targetted a second QSO population, not selected uniformly over the footprint, designed for clustering studies in the Ly-$\alpha$ forest. The Ly-$\alpha$ forest is a notably different kind of tracer of the matter density field than luminous galaxies. We therefore leave any identification of and analysis of voids in the Ly-$\alpha$ forest and in this second QSO population to future work.

\begin{table}[!ht]
\caption{Characteristics of eBOSS DR14 galaxy catalogues.} 
\centering
\begin{tabular}{lccc}
\hline\hline
& {$N_{obs}$} & {Effective area ($deg^2$)} & {$z$ range} \\
\hline
LRG  & 126656  & 2270 & $0.6 \leq z \leq 1.0$ \\
QSO  & 148750  & 2626 & $0.8\leq z \leq 2.2$ \\
\hline
\end{tabular}
\label{table:numberofgal}
\end{table}

\subsection{Mock galaxy catalogues}

In this section we describe the mock galaxy catalogues used in our analysis. The fiducial cosmology of the mocks is as follows: $\Omega_m=0.307115$ , $h=0.6777$, $\sigma_8=0.8225$, $\Omega_b=0.048206$, $n_s=0.9611$. We have used this cosmology throughout the paper. This is the same cosmology that has been used in other eBOSS DR14 papers.

\subsubsection{Quick Particle Mesh mock catalogues}
\label{sec:QPMmocks}

To construct our covariance matrix for the LRG sample, and to test our methodology, we used 499 Quick Particle Mesh \citep[QPM,][]{White2014} mock galaxy catalogues. A full description of the production of these catalogues can be found in ref. \citep{Alam2017}. QPM uses low resolution mesh simulations to evolve an initial density field. Particles were then sampled from the density field such that their one and two-point statistics match those of dark matter haloes. A halo-occupation distribution was applied to populate each simulation with galaxies and thus construct galaxy density fields that match the observed clustering of BOSS galaxies. 
Each mock is designed to match both the angular selection function of the survey, including effects such as fibre collisions, and the observed redshift distribution of the target tracer.

\subsubsection{Extended Zel'dovich mock catalogues}
\label{sec:EZmocks}

To emulate the QSO sample we used Extended Zel'dovich mock catalogues. These were  produced following the method described in Ref.  \citep{Chuang2015}. The method uses the Zel'dovich approximation, which is sufficently accurate on large scales to account for the three-dimensional cosmic web. However, the Zel'dovich approximation yields only a crude approximation of the dark matter density field on small scales. The method therefore adds simple prescriptions for scale dependent, non-local, and non-linear biasing contributions, and for physical effects such as tidal fields. We use a total of 499 EZ mocks in our study.

\section{Void finding in eBOSS}
\label{sec:voidfinding}

In this section we describe our search for voids in the tracer populations described in Section \ref{sec:data}, and present summary statistics of the subsequent void catalogues.

\subsection{The Void Identification and Examination toolkit}

The Void Identification and Examination toolkit (VIDE) is a toolkit for identifying voids and calculating key statistics \citep{Sutter2015VIDE}. At the heart of VIDE is the ZOBOV (ZOnes Bordering On Voidness) watershed void finder \citep{Neyrinck2008, LavauxWandelt2012, Platen2007}. ZOBOV first builds a Voronoi tessellation of the tracer particles and then assigns a density to each cell based on its Voronoi volume \citep{Platen2011}. The survey volume is then split into `zones' corresponding to attraction basins of the local density minima. Zones are then assembled into voids by joining zones that share the lowest common saddle point in the density field. It is worth noting that in practice, when analysing current data, voids are rarely composed of multiple zones. Thus VIDE defines voids in such a way that overdense structures such as clusters, filaments, and walls naturally divide a given volume into voids. This means that there is a good correspondence between what one perceives as a void by eye when looking at the cosmic web and what VIDE identifies as a void. This correspondence is clearly demonstrated in Figure 1 of \citep{Colberg2008}, where the performance of ZOBOV is compared to that of a number of other void finders. It should be noted that particles in surrounding walls are also included in the voids. In principle the ZOBOV algorithm is parameter free. 
In practice, however, following \cite{Neyrinck2008}, VIDE 
only joins zones if the separating ridge falls below $20\%$ of the mean tracer density. This procedure reduces the probability of identifying Poisson fluctuations as voids and prevents the full volume from being identified as a void. For a more detailed discussion on our efforts to 
identify and remove Poisson voids, please see Section \ref{sec:MVA}.

The centres of VIDE voids are defined as the volume-weighted barycentres of the ensemble of cells 
comprising each void. Other authors may define void centres in different ways, for a discussion on other possible centre definitions see, for example, Refs. \cite{NadathurHotchkiss2015, Nadathur2017}. VIDE does not assume any spherical symmetry, indeed watershed voids often have highly irregular shapes. Nevertheless, each void is characterised by an effective radius, $r_v$, defined as the radius of a sphere with the same volume as the sum of the Voronoi volumes, $V_j$, comprising the void,
\begin{equation}
r_v = \bigg(\frac{3}{4\pi}\sum_j V_j\bigg)^{1/3}.
\end{equation}
In much of what follows, separations and cross-correlations are expressed as functions of distance in units of $r_v$. This allows us to stack voids of different radius on top of one another easily. Without rescaling, overdensities around small voids can cancel out the signal from the underdense interiors of large voids, making the shape of the resulting mean profile difficult to interpret. We note that we also ran tests dividing our void catalogue into bins of different effective radius. 

\subsection{Taking into account survey geometry}

The eBOSS survey consists of two target volumes, separated in the sky by the plane of the Milky Way: the North Galactic Cap (NGC) and the South Galactic Cap (SGC). When running VIDE we combined the NGC and the SGC samples into a single volume so that only one run of the void finding algorithm was needed.

Void finding is affected by survey boundaries in both angular and redshift space. It is therefore necessary to construct an appropriate mask around the survey, so that these boundaries are well defined and so that the void finder does not search for voids outside of the survey. This was done by loading the data into a {\sc healpix} map of $N_{side} = 128$ to get a rough estimate of the survey boundaries ($N_{side}$ defines the number of divisions along the side of a base-resolution pixel and thus the resolution of the pixelisation). 

Placing `mock' particles around the survey boundaries prevents the VIDE algorithm from identifying regions outside of the survey as belonging to voids. To avoid the identification of spurious voids and to ensure that the void sample is stable to minor changes in the parameters of the algorithm, it is important that the number density of mock particles is as high as is computationally feasible. We tried several different values and found that a number of mock particles around five times the number of galaxies gave stable results, i.e. increasing the number of random mock particles beyond this had no affect on the output void catalogue. 
The number of mock particles therefore varied from sample to sample, being a function of the number of galaxies in the catalogue. As the final step, voids with mock particles within their effective radius were discarded. 
This pruning tempers the risk of finding truncated voids near the survey edges or spurious voids arising from the presence of mock particles. 

\subsection{Abundance as a function of effective void radius}
\begin{figure}[h!]
\includegraphics[width=.9\textwidth]{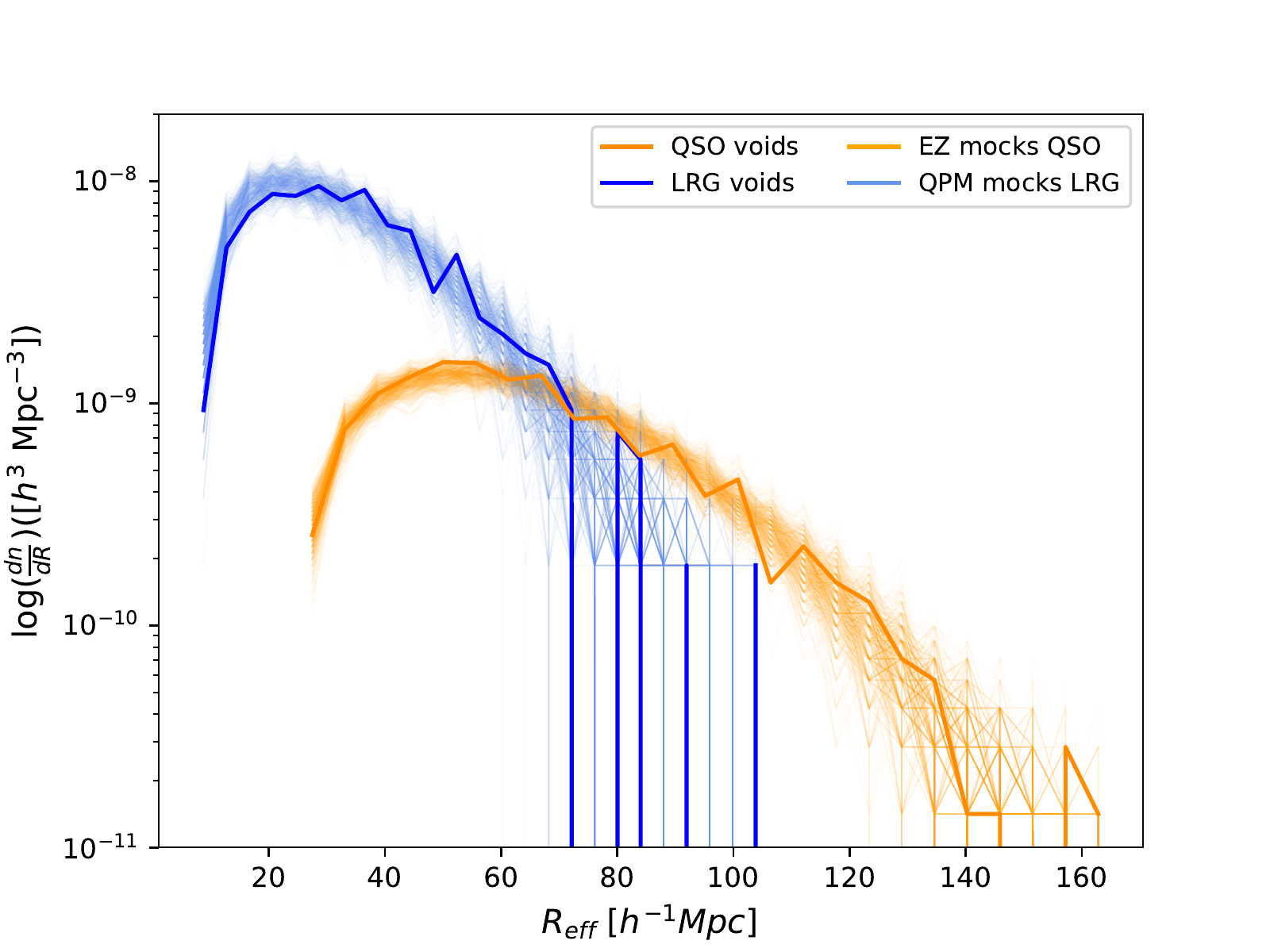}
\caption{\label{fig:abundance} Abundance of voids in the LRG sample (thick blue curve) and QSO sample (orange curve), as a function of effective radius. The thin blue and orange lines correspond to the abundance of voids in the QPM and EZ mocks, respectively.}
\end{figure}

The measurement and modelling of the abundance of voids as a function of their effective radius is an active and promising field of research \citep{Jennings2013, Ronconi2017, Sheth2004, Contarini2019, Verza2019}. The abundance of voids has been demonstrated to be a sensitive probe of the dark energy equation of state \citep{Pisani2015}, and of other cosmological parameters such as the neutrino mass \citep{Kreisch2018, Massara2015}.
Figure \ref{fig:abundance} shows the abundance of voids as a function of their effective radius, as measured in the LRG and QSO samples. Also shown are the abundances in the corresponding mock catalogues.  In this paper, we do not attempt to obtain any cosmological constraints from these abundance curves, leaving this for further work. However, the abundance curves are still very useful descriptive statistics for comparing voids in different tracer populations. The abundance of voids also serves as a test of the void finder; the abundance of voids in a Poisson sampling of a uniform random field looks quite different from that in a sampling of an evolved density field \citep{Cousinou2019}. 
The two curves both have the same basic shape, typical of void abundance curves, with larger voids being exponentially less common than smaller voids. The abundance of voids in the data shows a remarkable similarity to the abundance of voids in the mocks. Although we do not make any formal constraints here using the abundance, this agreement can be used to argue that the measured abundance rules out any large deviation from $\Lambda$CDM. %

Table \ref{table:stats} shows the number; minimum and maximum radius of voids; and the mean particle separation, $mps$, in the two samples.  
Since the LRG sample is denser and has a lower bias than the QSO sample, it is no surprise that LRG voids are smaller than QSO voids.
Likewise, the QSO sample is the most highly biased and the sparsest, it also covers the largest cosmological volume. The QSO voids are therefore, on average, much larger. The largest void in the QSO sample has a radius $\sim1.5$ times that of the largest void in the LRG sample.

\begin{table}[!ht]
\caption{Characteristics of voids identified in eBOSS DR14 galaxy and quasar catalogues, and in associated mock galaxy catalogues. The errors on the mock values come from the variance of the mocks.}
\centering
\begin{tabular}{rcccc}
\hline\hline
& {$N_v$} & {$r_{\rm min} (h^{-1}{\rm Mpc})$} & {$r_{\rm max} (h^{-1}{\rm Mpc})$} & mps  $(h^{-1}{\rm Mpc})$ \\
\hline
Data catalogues & & & & \\
LRG  & 471  & 6.78 & 105.82 & 25.73\\
QSO  & 970  & 24.67 & 165.65 & 45.06 \\
\hline
Mock catalogues & & & & \\
QPM LRGs & $482 \pm 1$ & $7.410 \pm 0.004$ & $99.7 \pm 0.4 $  & $ 25.950 \pm 0.002 $ \\
EZmock QSOs & $956 \pm 1$ & $23.34 \pm 0.07$ & $ 163.2 \pm 0.5 $  & $ 45.320 \pm 0.002 $ \\
\hline
\end{tabular}
\label{table:stats}
\end{table}

\subsection{Redshift distribution}

\begin{figure}[h!]
\includegraphics[width=.9\textwidth]{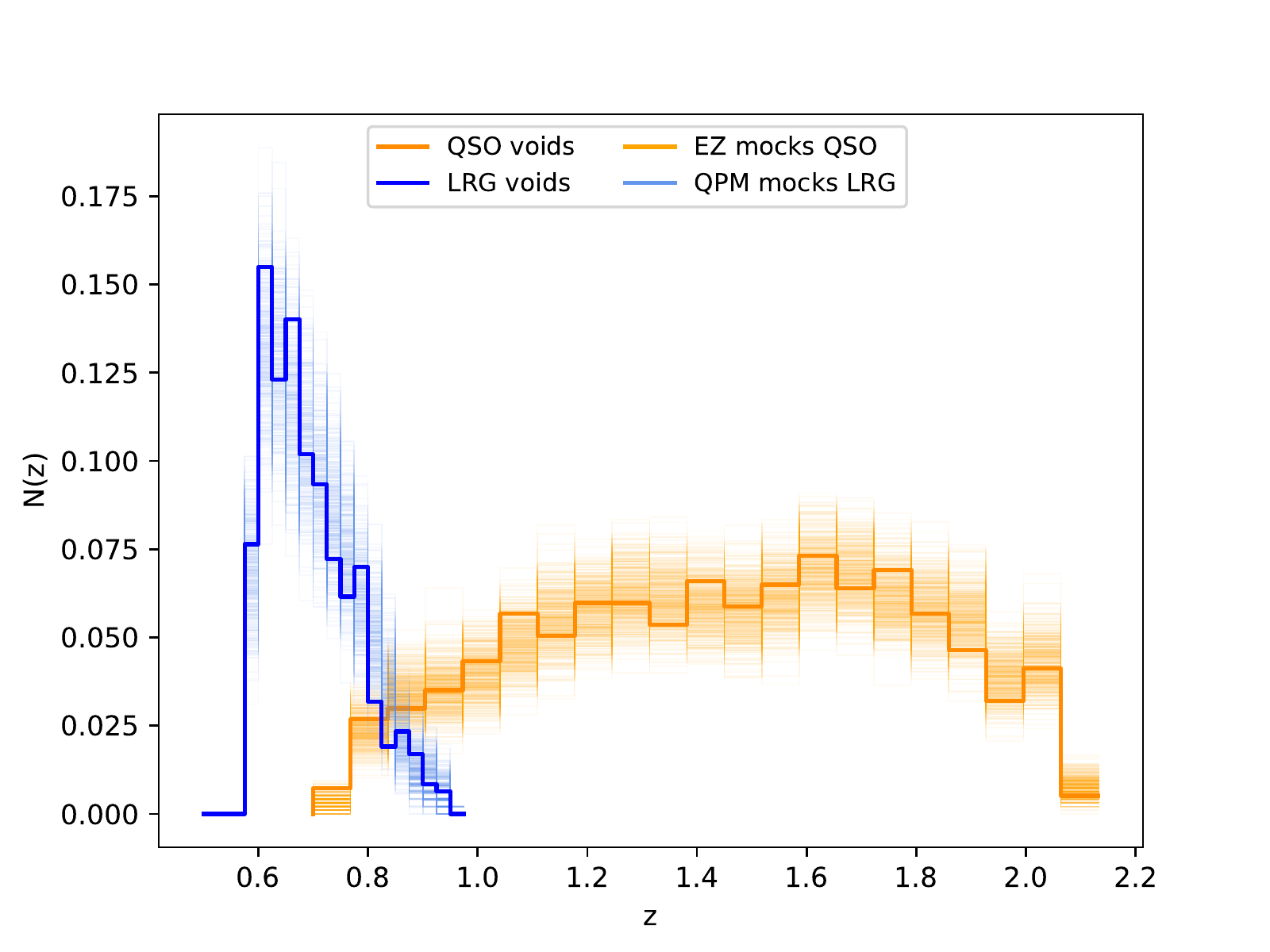}
\caption{\label{fig:VNz2} Histograms of the redshift distribution of voids, normalised so that the integral of the distribution is equal to unity. The thick blue curve shows the distribution of LRG 
voids, the thick orange curve indicates the distribution of QSO voids. The thin blue and orange curves show the redshift distributions of voids in the QPM and EZ mocks, respectively.}
\label{fig:voidzdist}
\end{figure}

Figure \ref{fig:voidzdist} illustrates the redshift distribution of voids in the two samples. The redshift distribution of voids identified using VIDE broadly follows that of the tracer population. The redshift distribution of QSO voids is particularly flat. Although the volume of the QSO sample increases with redshift, so does its sparsity. This effect is compounded by the fact that voids in the QSO distribution are generally large, due to the high bias of QSOs. Therefore a lower density of larger voids is identified at these high redshifts. A notable feature of the redshift distribution of voids is that the number of voids decreases close to the upper and lower redshift limits of the tracer sample. This bevhaviour is due to the fact that voids intersecting angular and redshift survey boundaries have been excised from the catalogue. These voids have been removed because parts of them may lie outside of the survey and thus the true size and shape of these voids is difficult to estimate.

\section{Measurement of the void-galaxy cross-correlation}
\label{sec:measurement}

The void-galaxy cross-correlation, $\xi_{vg}$, synonymous with the void density profile, $\delta_g(r) = \frac{n_g(r)}{\bar{n_g}} -1$, describes the average over (under) density of galaxies as a function of distance from the void centre. Here, we employ the Davis and Peebles estimator for the void-galaxy cross-correlation \citep{DavisPeebles1983},
\begin{equation}
\xi_{vg}(r) = \frac{D_vD_g}{D_vR_g} - 1,
\end{equation}
where $D_vD_g$ is the number of void-galaxy pairs with separation $r$, and $D_vR_g$ is the number of void-random pairs. The random catalogue follows the angular and redshift selection function of the galaxy sample in question. 
Thus the estimate of the correlation function takes into account any effects introduced by the mask. Some authors adopt the Landy-Szalay estimator to calculate the void-galaxy cross-correlation \citep{LandySzalay1993}. The Landy-Szalay estimator requires a separate random catalogue, $R_v$, which follows the void selection function. It also requires estimating the number of $R_vR_g$ and $D_gR_v$ pairs. Ref. \citep{Hamaus2017Multipoles} point out that the contribution of these factors to the monopole and quadrupole of the void-galaxy cross-correlation function is negligible. We therefore choose to employ the Davis \& Peebles estimator in this work.

\subsection{Redshift space distortion model}
\label{sec:RSDmodel}

The multipoles of the galaxy autocorrelation function have been used for some time to constrain RSD \citep{Hamilton1992}. Ref. \citep{Cai2016} were the first to demonstrate, using simulations, that the multipoles of the void-galaxy cross-correlation function can be used to measure the growth rate of structure. Here, we follow the method applied by ref. \cite{Hamaus2017Multipoles}, who were the first to measure the growth rate around voids in SDSS data using a multipole method. The efficacy of this methodology has subsequently been confirmed by ref. \citep{Achitouv2019}.

The anisotropic void-galaxy cross-correlation can be written as the sum of its multipoles,
\begin{equation}
\xi_{vg}^{s}(r,\mu) = \sum\mathcal{L}_\ell(\mu) \xi_\ell(r),
\end{equation}
where the superscript $s$ denotes redshift space and where
\begin{equation}
\xi_\ell(r) = \int_0^1 \xi_{vg}^{s}(r,\mu)(1+2\ell)\mathcal{L}_\ell(\mu){\rm d}\mu,
\end{equation}
where the Legendre polynomials, $\mathcal{L}_{\ell}(\mu)$, are defined as
\begin{eqnarray}
\mathcal{L}_0(\mu) &=& 1, \\
\mathcal{L}_2(\mu) &=& \frac{3\mu^2 - 1}{2}.
\end{eqnarray} 
The variables $r$ and $\mu$ are defined as
\begin{eqnarray}
r &=& \sqrt{r_\perp^2 + r_\parallel^2}, \\
\mu &=& \frac{r_\parallel}{\sqrt{r_\perp^2 + r_\parallel^2}}.
\end{eqnarray}
The distance $r$ is the separation from the void centre in units of the effective radius of the void, $r_v$. This convention of dividing by the void radius allows us to easily stack voids of different size together and to coherently capture the form of the average void density profile of the sample. The quantity $\mu$, is the cosine of the angle to the line-of-sight.

For the case that the velocity field around cosmic voids is strictly linear and the void centres are stationary, all multipoles higher than the quadrupole, and all odd multipoles, vanish. It has been shown that the centres of voids do indeed move over time \cite{Chuang2015, Massara2018}, and some authors have argued that reconstruction of the real-space positions of tracers is necessary in order to identify voids \cite{Nadathur2018}. However, Ref. \cite{Massara2018} reported that on average voids only move a few $h^{-1}$Mpc over their lifetimes. They demonstrate that void motions do not appear to be a function of the size of voids (see Figure 8 of \cite{Massara2018}). These motions are therefore more important for smaller voids. Most of the voids we study in this paper are larger than those studied in \cite{Massara2018}. Thus for this paper we stick with the assumption that void centres are stationary. Following the model introduced in \cite{Cai2016}, the relationship between the non-vanishing multipoles and the undistorted density profile can be written,
\begin{eqnarray}
\label{eqn:monopole}
\xi_0(r) &=& \bigg(1+\frac{\beta}{3}\bigg)\xi_{vg}(r),\\
\xi_2(r) &=& \frac{2\beta}{3}[\xi_{vg}(r) - \bar{\xi}(r)],
\label{eqn:quadrupole}
\end{eqnarray}
where $\bar{\xi}(r)$ is the cumulate average correlation function defined by, 
\begin{equation}
\label{eqn:xibar}
\bar{\xi}(r) = \frac{3}{r^3}\int_0^r \xi(r')r'^2 {\rm d}r'.
\end{equation}

Thus a measurement of $\beta$ can be obtained via a comparison of the multipoles,
\begin{equation}
\xi_0(r) - \bar{\xi}_0(r) = \xi_2(r)\frac{3+\beta}{2\beta},
\label{eq:model}
\end{equation}
where $\bar{\xi}_0(r)$ is the cumulate average monopole calculated by substituting the monopole into eq. \ref{eqn:xibar}. All the quantities in Eq. \ref{eq:model} are measured from data and so there is no need to
assume a density profile. The growth rate of structure, $\beta$, can be measured by minimising the function,
\begin{equation}
\epsilon_i = \frac{2\beta}{3+\beta} \big[ \; \xi_0(r_i) - \bar{\xi}_0(r_i) \; \big] - \xi_2(r_i),
\end{equation}
where in an ideal scenario, $\epsilon_i(\beta_{\rm true})=0$, in each radial bin, $r_i$. The best fitting value of $\beta$ is then the value which maximises the likelihood,
\begin{equation}
\mathcal{L}(\xi_0 | \beta) = \frac{1}{(2\pi)^{N/2}\sqrt{{\rm det}\mathcal{C}}} \exp \bigg(-\frac{1}{2}\mathcal{E} \mathcal{C}^{-1} \mathcal{E}^T \bigg),
\end{equation}
where $\mathcal{E}$ is the vector of residuals $\epsilon_i$ in each radial bin, $N$ the number of bins, and $\mathcal{C}$ is the covariance matrix, $\mathcal{C}_{ij} = \langle\epsilon_i\epsilon_j\rangle$. The normalisation factor, $\sqrt{{\rm det}\mathcal{C}}$, is important because the covariance matrix is a function of the parameter we are trying to estimate, $\beta$ \cite{Hamaus2017Multipoles}. We estimated the covariance matrix for each void sample using 499 mock galaxy catalogues (a description of the mock catalogues can be found in sections \ref{sec:QPMmocks} and \ref{sec:EZmocks}) via,
\begin{equation}
C_{ij} = \frac{1}{N_m - 1}\sum_{k=1}^{N_m}(\epsilon_i^k - \tilde{\epsilon}_i)(\epsilon_j^k - \tilde{\epsilon}_j),
\end{equation}
where $\epsilon_i^{k}$ is the residual in radius bin $r_i$ for the $k^{\rm th}$ mock galaxy catalogue, and $\tilde{\epsilon}_i$ is the mean residual in that bin considering all of the mocks. 

\begin{figure}
    \centering
    \begin{subfigure}[t]{0.45\textwidth}
        \centering
        \includegraphics[width=\linewidth]{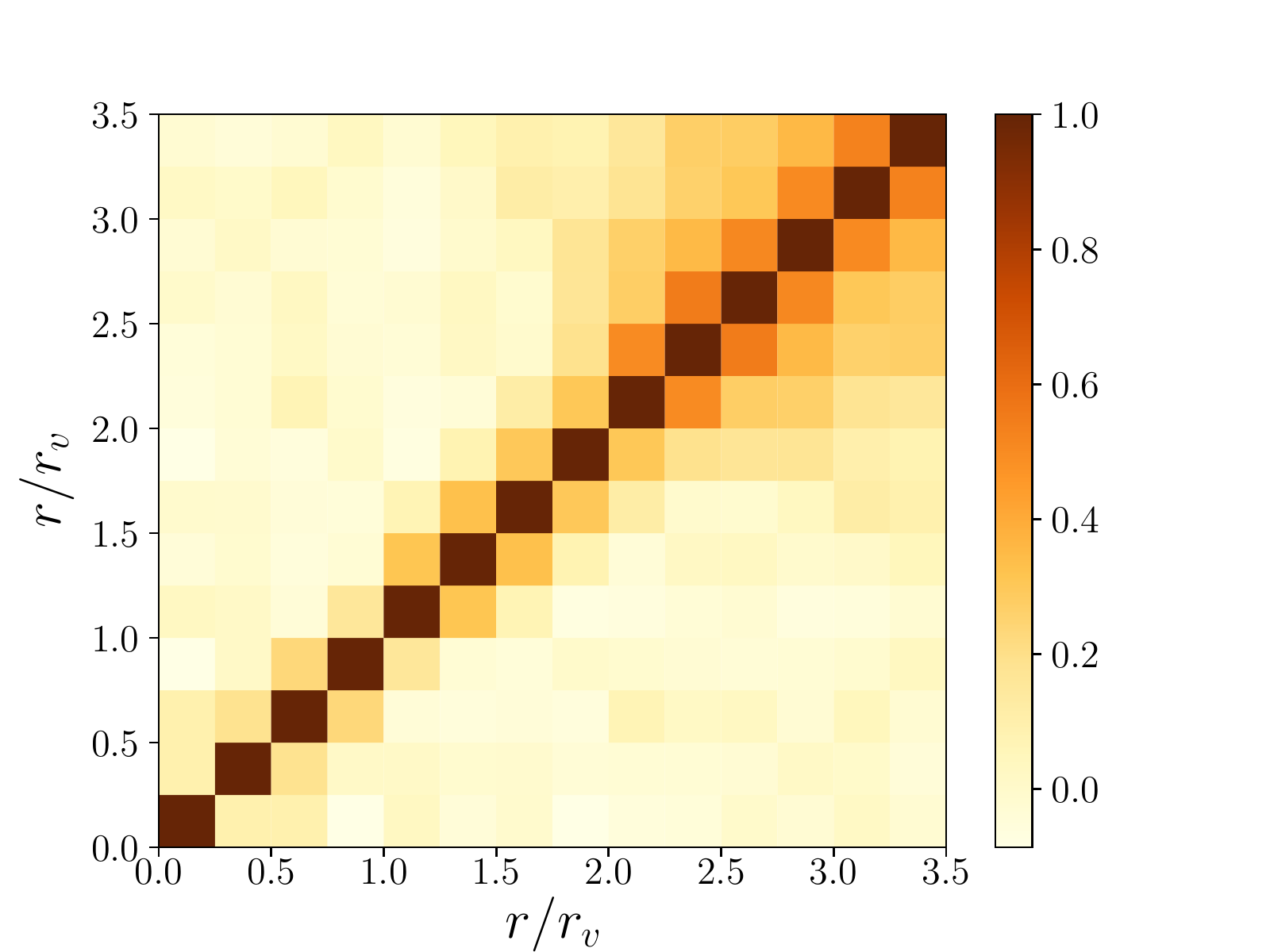} 
        \caption{} 
        \label{fig:cormatLRG}
    \end{subfigure}
    \hfill
    \begin{subfigure}[t]{0.45\textwidth}
        \centering
        \includegraphics[width=\linewidth]{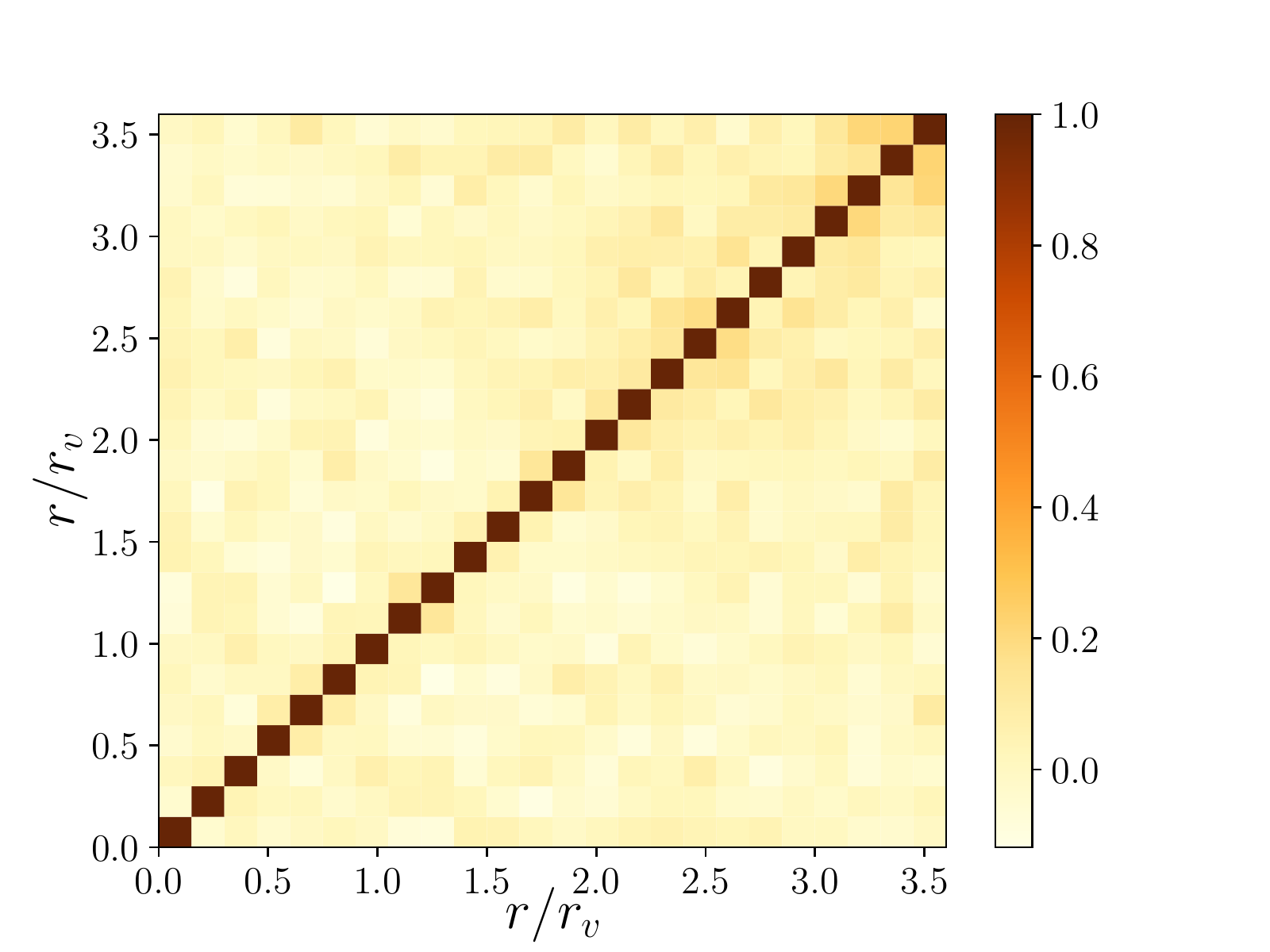} 
        \caption{} 
        \label{fig:cormatQSO}
    \end{subfigure}
    \caption{Correlation matrix produced from 499 QPM mock catalogues, left subpanel (a), and from 499 EZ mock catalogues, right subpanel (b).}
\end{figure}

Figures \ref{fig:cormatLRG} and \ref{fig:cormatQSO} show the correlation matrices produced using the LRG-like QPM mocks and the QSO-like EZ mocks, respectively. One can see that the off-diagonal components are not extremely strong, though they are more apparent in the covariance matrix from the LRG-like mocks than from the QSO-like mocks. These off diagonal components are most prominent at large void-galaxy separations.

\subsection{Tests on bin size}
The number of bins in void radius used to measure the multipoles can have an effect on the goodness of the fit of the model to the data. The more bins that are used the more detail can be captured when measuring and fitting the monopole. However, with a decrease in the width of each bin comes an increase in noise as fewer void-galaxy pairs have a separation that falls within a given bin. Reducing the number of bins decreases the noise levels but also smooths out detail in the multipoles. There is thus a trade-off between these two effects. The optimal number of bins can be found by measuring the $\chi^2$ per degree of freedom and choosing the number of bins that corresponds to a value closest to unity. A $\chi^2$ per degree of freedom much greater than unity implies a poor fit, whilst $\chi^2$ much less than unity implies over fitting.

Table \ref{tab:valbetaLRG} presents the value of the growth rate parameter $\beta$ recovered from the LRG void catalogue using different numbers of bins in $r/r_v$. We adopted 12 bins of width $\Delta(r/r_v) = 0.30$, as this option yeilds a reduced $\chi^2_N = 0.99$, which is closer to unity than the other bin widths tested. Table \ref{tab:valbetaQSO} lists the same information for the QSO void catalogue. For the QSO cross-correlation we chose to use 24 bins of width $\Delta(r/r_v) = 0.15$, which gives a reduced $\chi^2_N = 1.36$. 

\renewcommand{\arraystretch}{1.2}
\begin{table*}
   \caption{Values of $\beta$ for different $r/r_{v}$ bin sizes for the multipoles of the LRG void-galaxy cross-correlation, after the MVA cuts have been applied.} 
        \begin{center}
          \begin{tabular}{cccc}
            \hline\hline
             $\Delta(r/r_{v}) $ & bins & $\chi2_{N}^{part}$ & $\beta$ \\
             \hline
             0.20 & 18  & 1.55 &  $0.66_{-0.27}^{+0.31}$ \\
             0.25 & 14 & 1.33 &  $0.48_{-0.26}^{+0.31}$ \\
             0.30 & 12  & 0.99 &  $0.58_{-0.28}^{+0.33}$ \\
             0.35 & 10 & 0.68 &  $0.62_{-0.29}^{+0.35}$ \\
             0.40 & 9  & 0.64 &  $0.76_{-0.34}^{+0.43}$ \\
           \hline
        \end{tabular}
       \end{center}
   \label{tab:valbetaLRG}
\end{table*}

\begin{table*}
   \caption{Values of $\beta$ for different $r/r_{v}$ bin sizes for the multipoles of the QSO void-galaxy cross-correlation.} 
        \begin{center}
          \begin{tabular}{cccc}
             \hline\hline
             $\Delta(r/r_{v}) $ & bins & $\chi2_{N}^{part}$ & $\beta$ \\
             \hline
             0.10 & 36  & 1.69 &  $ 0.17_{-0.12}^{+0.12}$ \\
             0.12 & 30  & 1.86 &  $ 0.20_{-0.12}^{+0.14}$ \\
             0.15 & 24  & 1.36 &  $ 0.15_{-0.12}^{+0.13}$ \\
             0.18 & 20  & 1.59 &  $ 0.18_{-0.12}^{+0.14}$ \\
             0.20 & 18  & 1.75 &  $ 0.16_{-0.13}^{+0.14}$ \\
             0.25 & 14  & 1.83 &  $ 0.20_{-0.13}^{+0.15}$ \\
             0.30 & 12  & 1.90 &  $ 0.15_{-0.14}^{+0.15}$ \\
             \hline
        \end{tabular}
       \end{center}
   \label{tab:valbetaQSO}
\end{table*}

\subsection{Deep interiors}

The deep interiors of voids are often affected by the sparsity of tracers in these regions. Furthermore, when the central density of the void approaches $\delta_c\sim-1$, the relationship between the matter and velocity fields deviates from the simple linear model. Refs. \citep{NadathurPercival2017,Nadathur2018} argue that even when considering a strictly linear model, $\xi_0\sim-1$ implies that the series expansion used to derive equations \ref{eqn:monopole} and \ref{eqn:quadrupole} may not be sufficiently extended and thus further terms may need to be included in the expansion. This means that both the measurement and the modeling of the multipoles may be more difficult and less reliable close to the centre of voids. Ref. \citep{Cai2016} and ref. \citep{Achitouv6dF} mitigate this issue by masking out the central regions in their analyses. Both authors use void finders that work on the assumption of spherical symmetry. As a result, the central density of the voids they find is often close to $\delta_c\approx-1$. VIDE, on the other hand, finds voids which are less under-dense in their interiors. As a result, it is possible to measure the void-galaxy cross-correlation deep into the interior of VIDE voids. Therefore, these regions do not have to be masked and we can still assume linear theory.

The linearity of the relationship between velocity and density fields inside voids remains an active area of research. We plan to test modifications to the modelling of this relationship in future studies using eBOSS data. SDSS DR16 will cover a larger volume, so will provide us with more voids and greater statistical power. It will thus be more suitable for these studies. 

\subsection{Comparison with mock catalogues}

\begin{figure}
    \centering
    \begin{subfigure}[t]{0.49\textwidth}
        \centering
        \includegraphics[width=\linewidth]{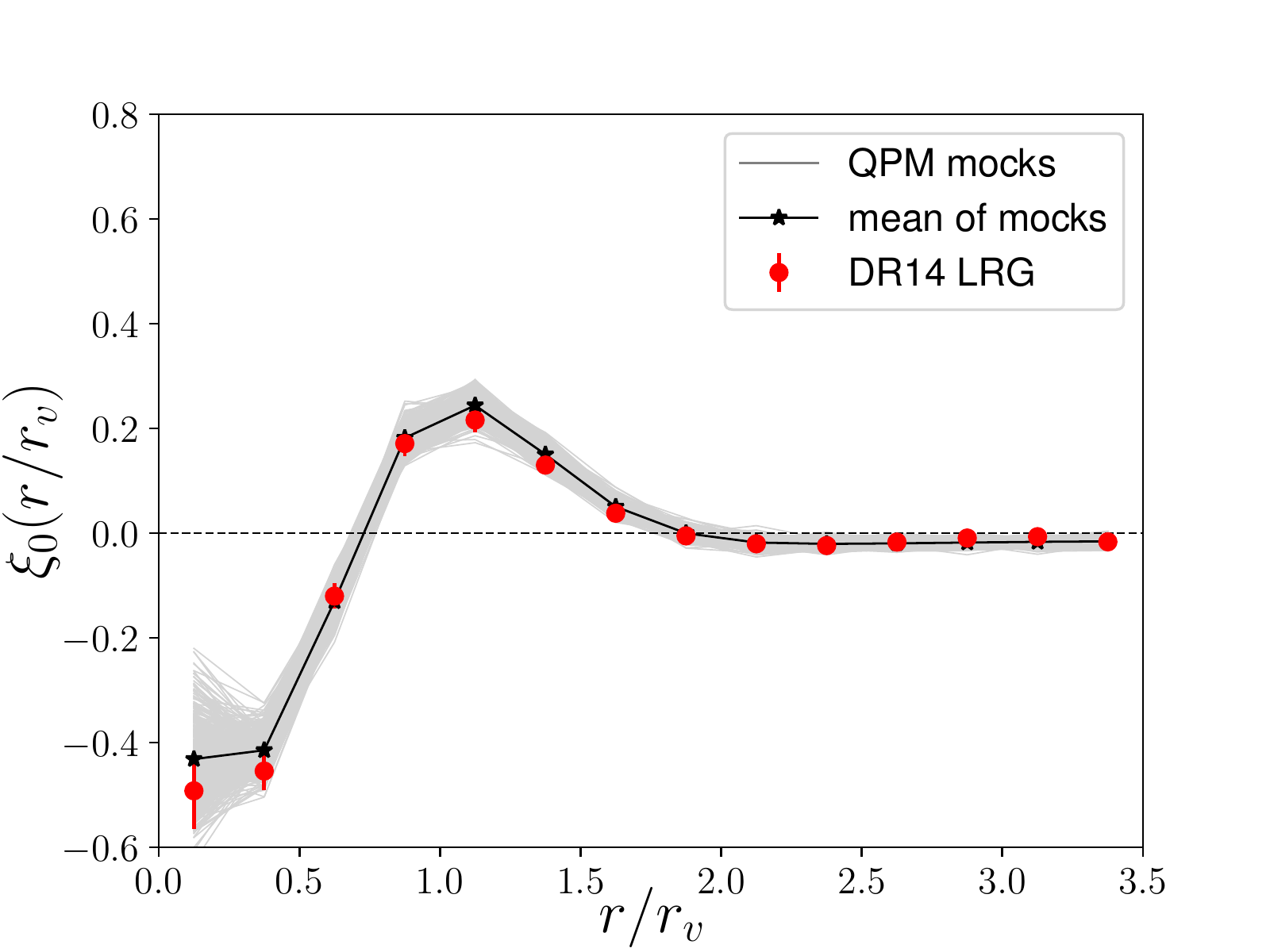} 
        \caption{LRG monopole} \label{fig:timing1}
    \end{subfigure}
    \hfill
    \begin{subfigure}[t]{0.49\textwidth}
        \centering
        \includegraphics[width=\linewidth]{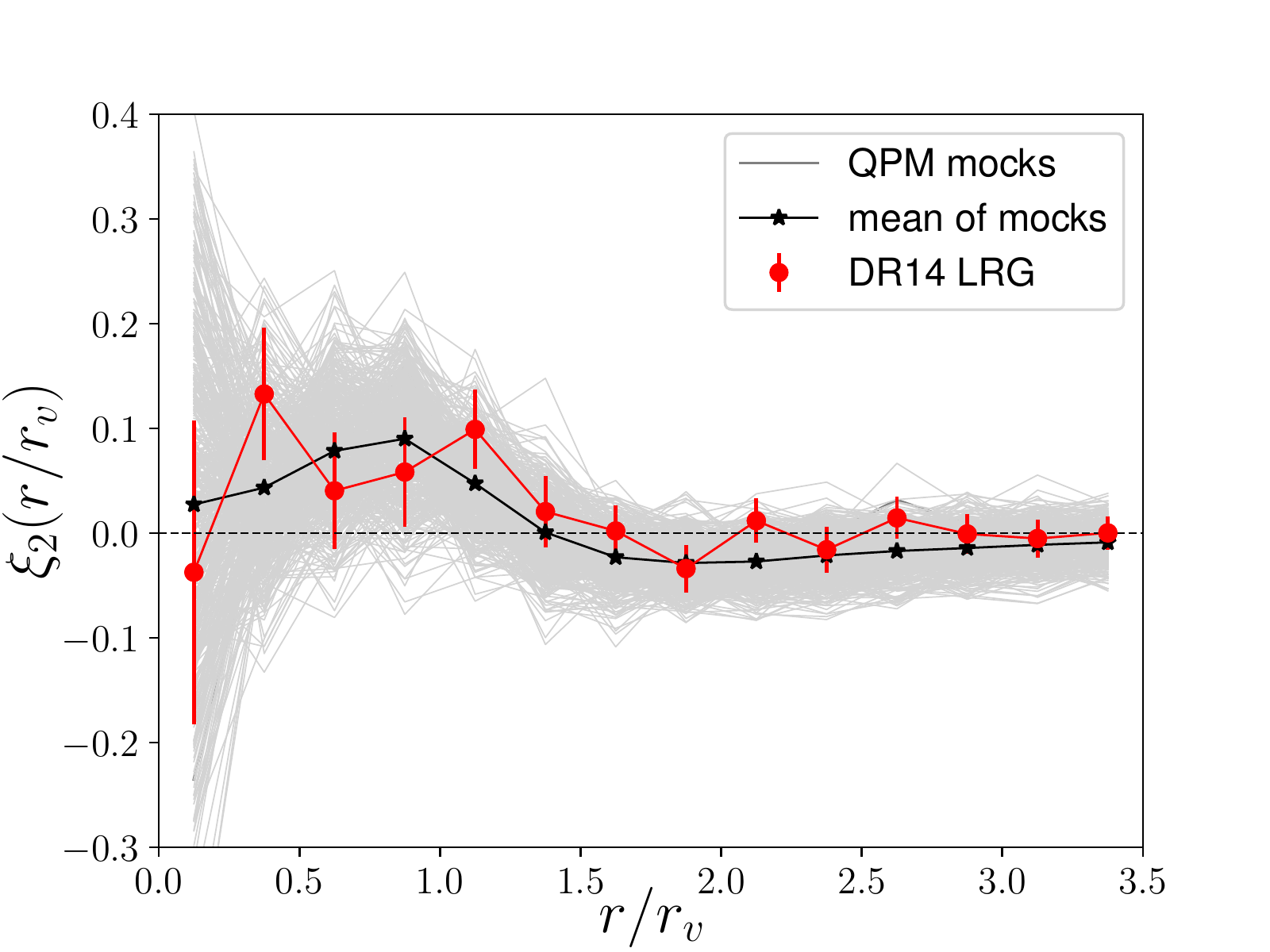} 
        \caption{LRG quadrupole} \label{fig:timing2}
    \end{subfigure}

    \vspace{1cm}
    \begin{subfigure}[t]{0.49\textwidth}
        \centering
        \includegraphics[width=\linewidth]{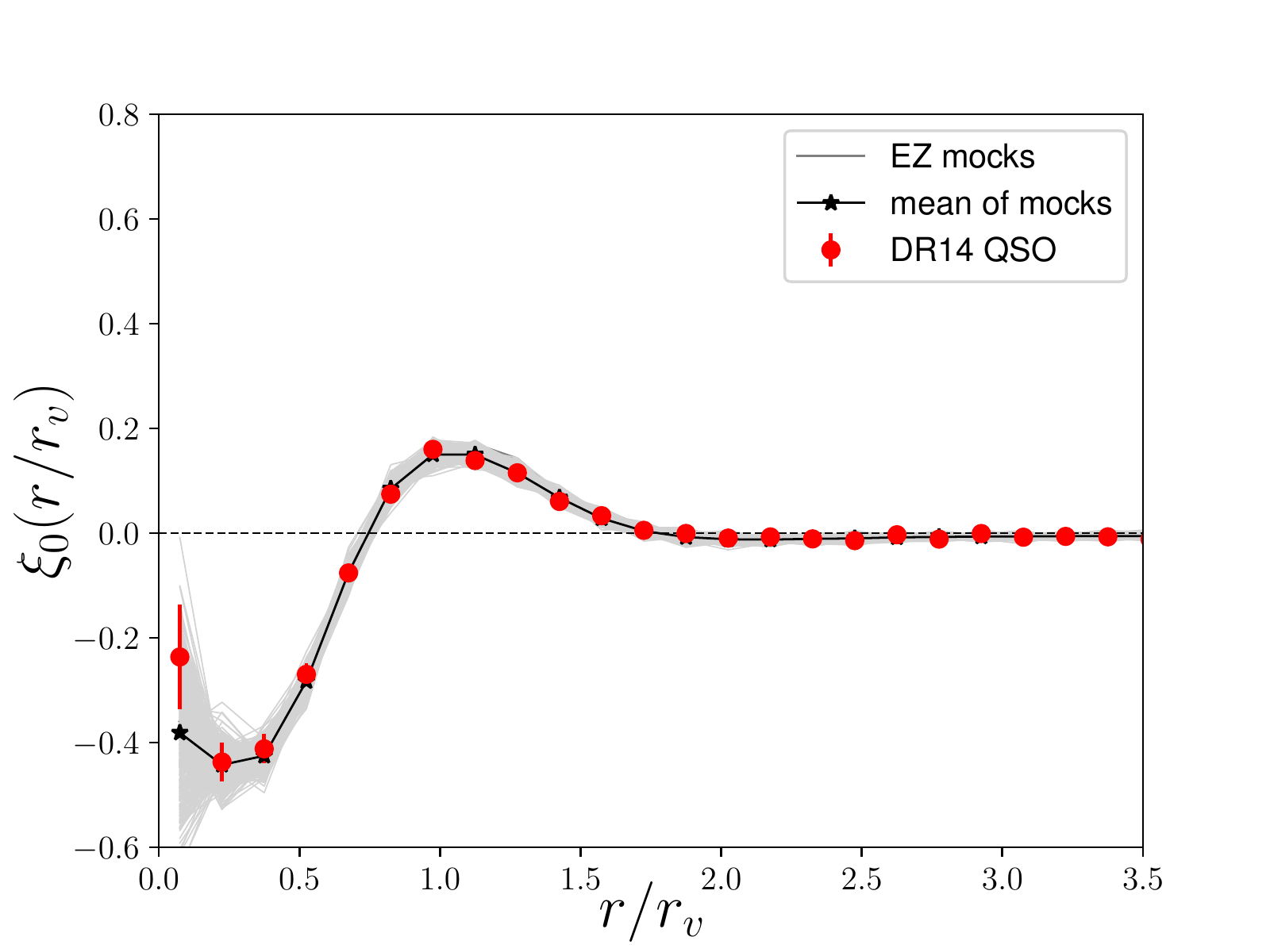} 
        \caption{QSO monopole} \label{fig:timing3}
    \end{subfigure}
    \hfill
    \begin{subfigure}[t]{0.49\textwidth}
        \centering
        \includegraphics[width=\linewidth]{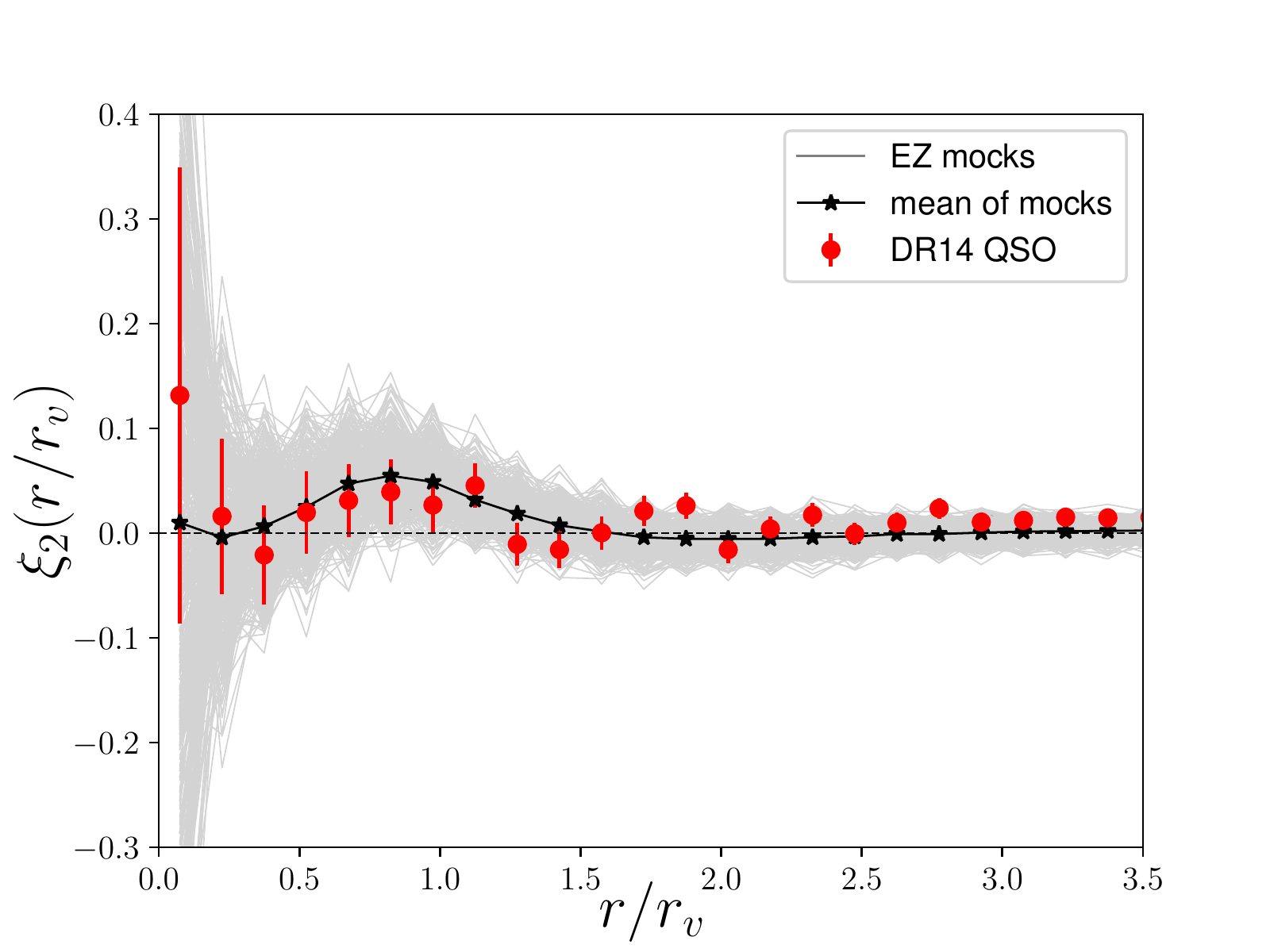} 
        \caption{QSO quadrupole} \label{fig:timing4}
    \end{subfigure}
    \caption{Comparison of multipoles of the void-galaxy cross-correlation function as measured in the data (red) and in mock catalogues (grey). The black line represents the mean value of the mocks. The error bars shown here are estimated from the diagonal of the covariance matrices.}
    \label{fig:mocks}
\end{figure}

Figure \ref{fig:mocks} shows a comparison between measurements of the monopole and quadrupole of the void-galaxy cross-correlation function in the data and the mocks, for the LRG and QSO samples. Also indicated in this figure is the mean of the mocks. For both samples, the monopole of the cross-correlation is precisely measured (the error bars are smaller than the points in many bins) and in very good agreement with the mocks. The quadrupole, in both cases, is noisier but does not show any strong deviation from the mocks.

\section{Multivariate analysis}
\label{sec:MVA}
To date, systematic biases and uncertainties on growth rate measurements from voids have largely been overlooked, because it has been assumed that they are sub-dominant to the statistical error. However, new data from forthcoming surveys will massively decrease the statistical error, therefore motivating the need to study systematic effects properly. This improvement will be essential if we wish to perform precision cosmology with voids in future surveys.

A known systematic affecting the measured value of the growth rate of structure using voids is the occurrence of spurious underdensities caused by the sparse sampling of galaxies. These `Poissonian' voids do not represent genuine underdensities in the cosmic web. Because they are not related to the matter density field there is no correlation between the peculiar velocities of galaxies and the location of Poissonian voids. Thus the expected quadrupole of these voids is null. They do, however, have a density profile, and thus a monopole, that exhibits some of the same features as genuine voids; even random catalogues have some regions that are emptier than others. For example, they are most underdense in their centres and at large separations the cross-correlation will tend towards the mean cosmic density. The presence of poissonian voids in a sample therefore dilutes the measured quadrupole signal and can also affect the measured monopole.

Ref. \citep{Cousinou2019} introduced the use of multivariate analysis techniques to remove spurious voids. The goal is to train a machine learning algorithm to distinguish between the two populations based on a number of input variables, in our case these are provided by the outputs of our void finding algorithm, VIDE. We do not know {\it a priori} which voids in a data or mock catalogue are true voids and which are Poissonian. However, by searching for voids in random catalogues, we are able to learn what the Poissonian voids look like. The problem is therefore one of separating the signal from a signal and background mixture by modelling just the background. The multivariate analysis algorithms are learning to identify the noise, i.e. the poissonian or spurious voids, and the assumption is that what is left is the signal. Following ref. \citep{Cousinou2019} we used two different supervised machine learning algorithms:
\begin{itemize}
    \item A {\bf Multi-Layer Perceptron} (MLP) is an artificial neural network that consists of at least three layers of nodes: an input layer, a hidden layer, and an output layer. An MLP is trained using supervised back propagation. The architecture of our MLP has one hidden layer of 11 nodes and uses the hyberbolic tangent activation function. The output is a number, between 0 and 1, that we call the MLP response.
    \item A {\bf Boosted Decision Tree} (BDT) is an extended cut based selection. All events are sorted by each input variable and then for each variable a value is found that best splits the sample into signal and background. This forms a branch with events falling either side of the criterion. The algorithm is repeated recursively on each branch and iterated until some stopping criterion is reached. The decision tree is boosted incrementally by training each new instance to emphasise training events that were mismodelled on previous iterations. The output, which we call the BDT response, is a number between -1 and 1.
\end{itemize}

For both algorithms we used five parameters output by the VIDE algorithm as input variables, namely: the normalised volume, {\it volnorm}; the density ratio, $\rho_{\rm cont}$; the core density, $\delta_{\rm core}$; the void probability, {\it voidprob}; and the number of particles defining the void, {\it numpart}. The redshift of the void is also an output of VIDE; however, preliminary analysis showed that it does not carry useful information for training our algorithms. For more information on the sensitivity of MVA techniques to various void features, please see \cite{Cousinou2019}.

For the LRG catalogue, the machine learning algorithms were trained using a signal training set built by running VIDE on 50 eBOSS-like QPM mocks, containing a total of 24\;368 voids. The background training set was built by running VIDE on 50 random catalogues, with each random catalogue containing 135\;000 particles. A total of 43\;628 voids were identified in the random catalogues. We used a BDT with 700 trees and trained the MLP with 600 back propagation cycles.

For the QSO void catalogue the MVA algorithms were trained using a data training set produced from 48 EZ mock galaxy catalogues, containing a total of 45\;680 voids. The background training set was produced using 49 random catalogues of 146\;000 particles each, containing a total of 53\;542 Poissonian voids. We used a BDT with 1000 trees and trained the MLP with 1000 back propagation cycles.

\subsection{Learning curves}

Learning curves give us an opportunity to diagnose bias and variance in supervised learning models. A Receiver Operating Characteristic (ROC) curve illustrates the diagnostic ability of a binary classifier system as its discrimination threshold is varied. In our case, the ROC curve shows the fraction of Poisson voids rejected (background rejection) as a function of the fraction of true voids retained (signal efficiency).  The ROC curve allows us to quantify the trade off between unwanted Poisson voids retained in the catalogue (false positives) and desired true voids thrown out (false negatives). The closer the curve is to the top right hand corner of the figure, the more true voids are retained and the more Poisson voids are rejected. 

\begin{figure}
    \centering
    \begin{subfigure}[t]{0.49\textwidth}
        \centering
        \includegraphics[width=\linewidth]{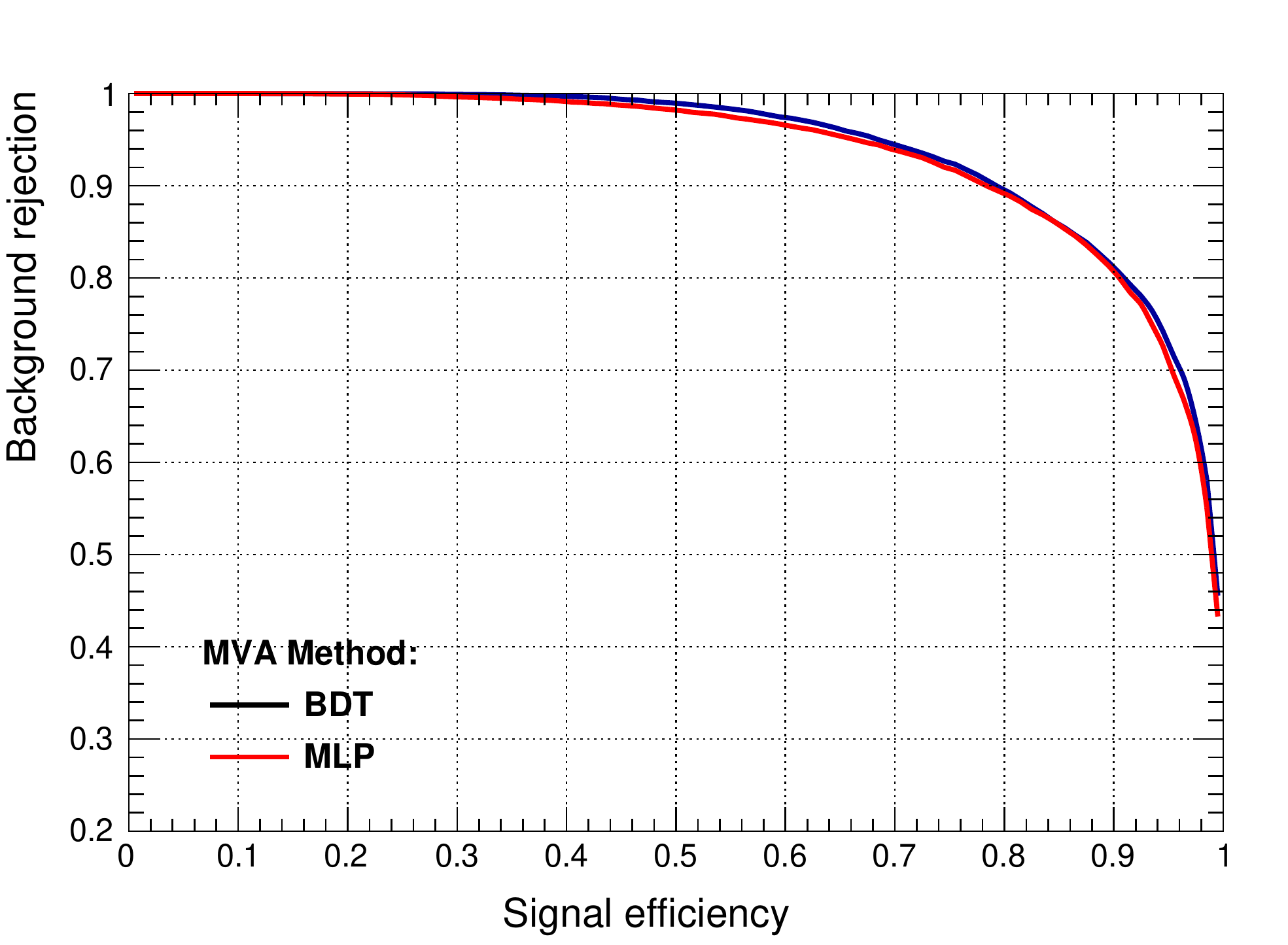} 
        \caption{LRG} \label{fig:LRGROC}
    \end{subfigure}
    \hfill
    \begin{subfigure}[t]{0.49\textwidth}
        \centering
        \includegraphics[width=\linewidth]{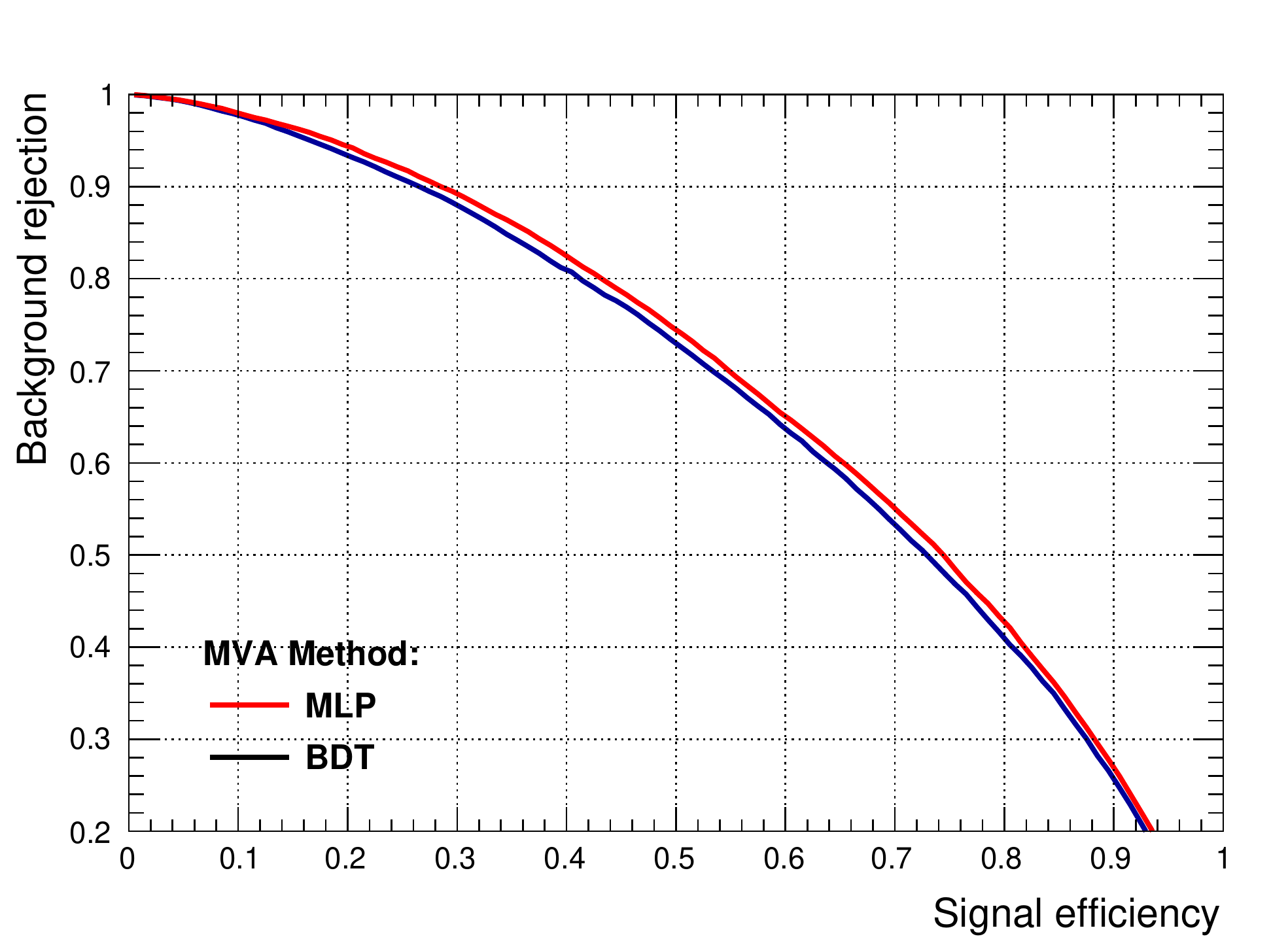} 
        \caption{QSO} \label{fig:QSOROC}
    \end{subfigure}
    \caption{ROC curves for the LRG voids sample, panel (a), and for the QSO voids sample, panel (b), from the MLP algorithm (red) and BDT algorithm (blue).}
\end{figure}

\begin{figure*}
\centering
    \begin{subfigure}[t]{0.49\textwidth}
    \centering
        \includegraphics[width=\columnwidth]{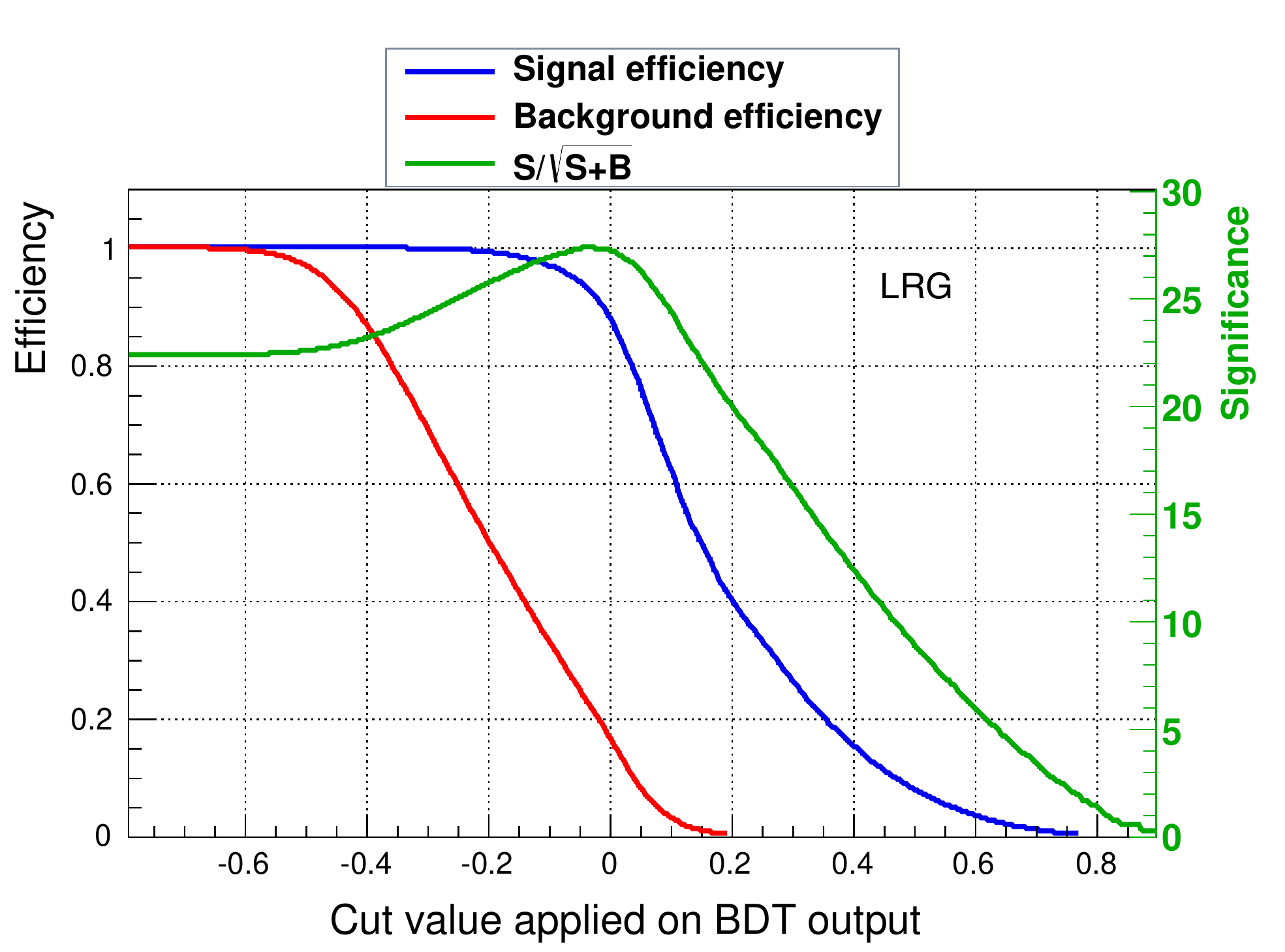}
    \caption{LRG}
    \label{fig:MVAeff}
       \end{subfigure}
    \begin{subfigure}[t]{0.49\textwidth}
    \centering
    \includegraphics[width=\columnwidth]{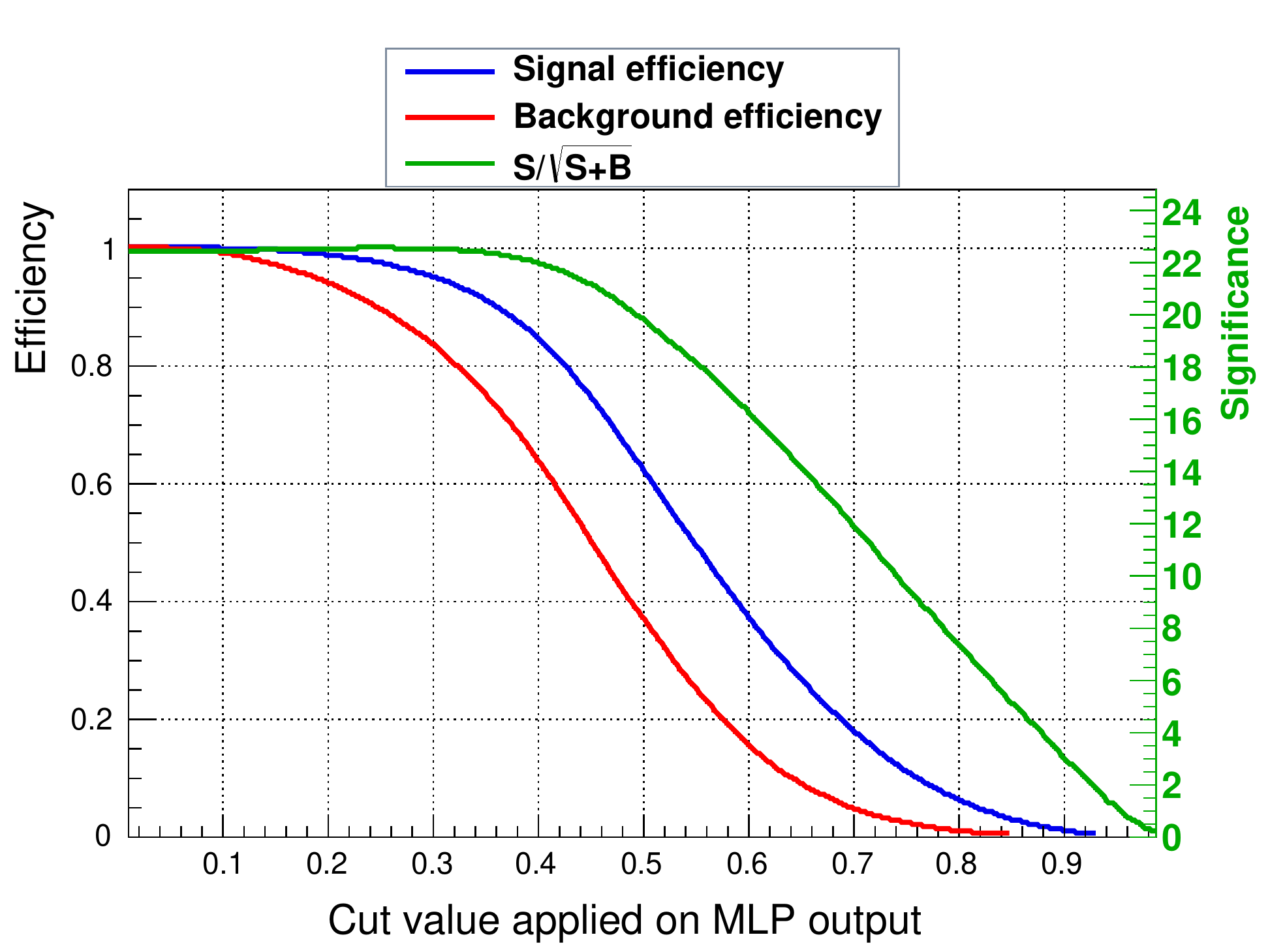}
    \caption{QSO}
    \label{fig:MVAeffQSO}
       \end{subfigure}
    \caption{The efficiency of the BDT classifier applied to the LRG voids, panel (a), and the efficiency of the MLP classifier applied to the QSO voids, panel (b). The optimal value on which to place the cut corresponds to the maximum of the green curve.}
       
\end{figure*}

For the LRG sample the ROC curve (Figure \ref{fig:LRGROC}) suggests that by using either the MLP or BDT algorithm it is possible to obtain a void sample that is both relatively pure and relatively complete. The difference between the discriminating power of these algorithms for the LRG data set is marginal. However, BDT slightly outperforms MLP and so we decided to use the BDT discriminator to cut our sample. We chose the value of $BDT\_response \geq  -0.04$, which corresponds to the maximum signal significance as implied by the green curve in Figure \ref{fig:MVAeff}. This gives us a catalogue of 419 LRG voids.

The ROC curves for the QSO sample are shown in Figure \ref{fig:QSOROC}. It is clear that both algorithms have more trouble separating Poisson voids from true voids in the QSO sample, with MLP slightly out performing BDT. It is more difficult to obtain a sample of QSO voids that has both a high purity and high completeness. The maximum of the green curve in Figure \ref{fig:MVAeffQSO} is not very clear and is close to zero. This plot indicates that it is not possible to excise the Poisson voids from the sample without also discarding almost all of the true voids. For this reason we decided not to impose a cut on the MLP output. The main benefit of the MVA in this case to demonstrate that for a sparse sample such as the QSOs the void catalogue is likely to be highly contaminated by Poisson voids that are difficult to remove. Thus the QSO void dataset may be unreliable.

\section{Results}
\label{sec:results}

\begin{table}[ht]
    \caption{Measurement of the growth rate in the eBOSS DR14 LRG and QSO samples.}
    \centering
    \begin{tabular}{lcccc}
        \hline\hline
        & {$\bar{z}$} & {$b_g$} & {$\beta$} & {$f\sigma_8$} \\ 
        \hline
        LRG & 0.703 & $2.30 \pm 0.03$ \cite{Zhai2017} & $0.58^{+0.33}_{-0.28}$ & $0.76^{+0.43}_{-0.37}$                                                        \\
        QSO & 1.53  & $2.45 \pm 0.05$ \cite{Laurent2017} & $0.15^{+0.13}_{-0.12}$ & $0.14^{+0.12}_{-0.11}$ \\
        \hline
    \end{tabular}
    \label{table:stats2}
\end{table}

The left hand panels of Figure \ref{fig:mocks} show the monopoles of the LRG voids (top left, panel a) and the QSO voids (bottom left, panel c). In both cases the monopole clearly shows the features characteristic of the void density profile, i.e. an underdense interior and an overdense ridge surrounding the void. The monopoles also show strong agreement with the monopoles measured in the mock galaxy catalogues.

\begin{figure}
    \centering
    \begin{subfigure}[t]{0.45\textwidth}
        \centering
        \includegraphics[width=\linewidth]{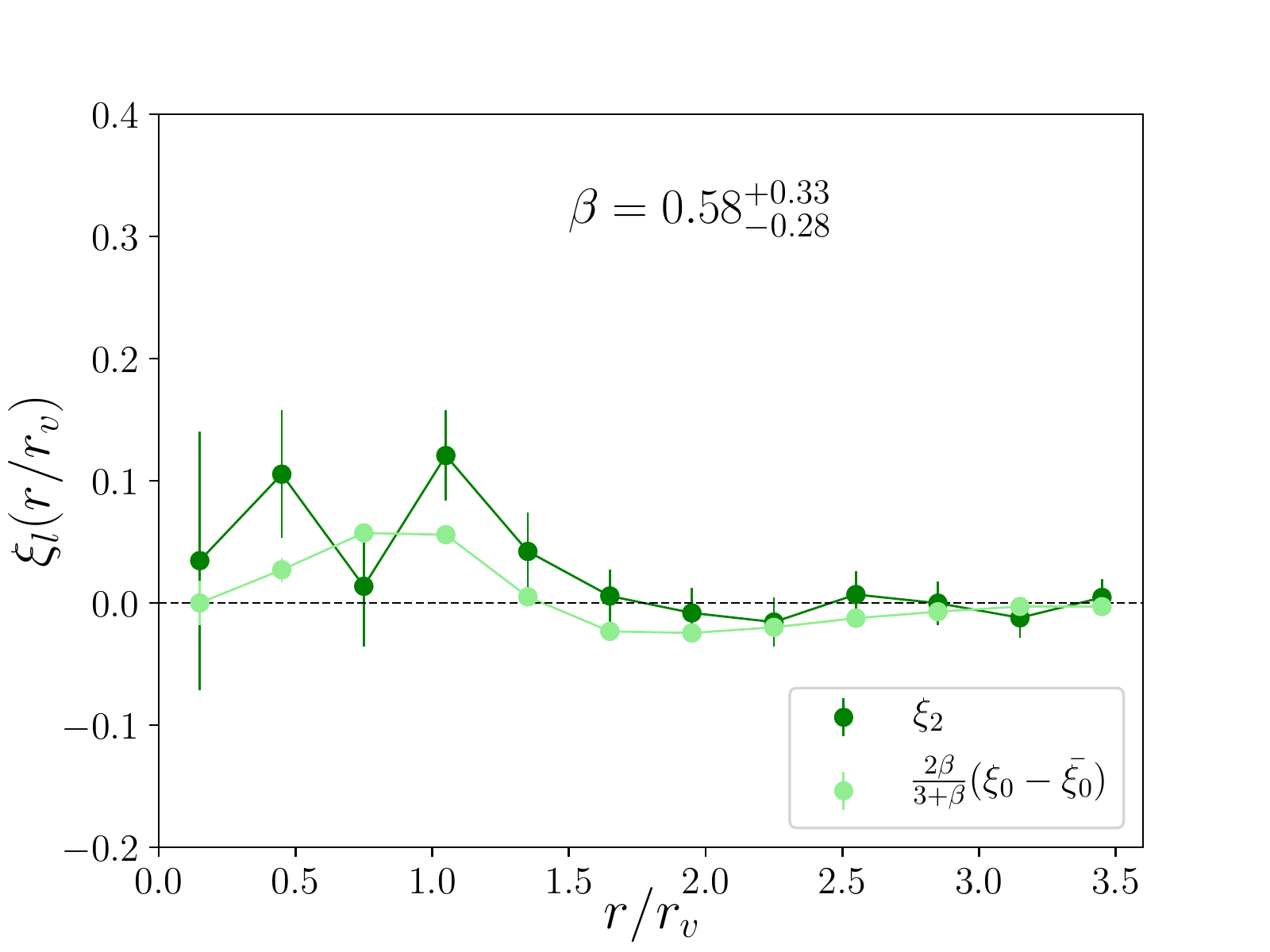} 
        \caption{LRG} 
        \label{fig:LRGbeta}
    \end{subfigure}
    \hfill
    \begin{subfigure}[t]{0.45\textwidth}
        \centering
        \includegraphics[width=\linewidth]{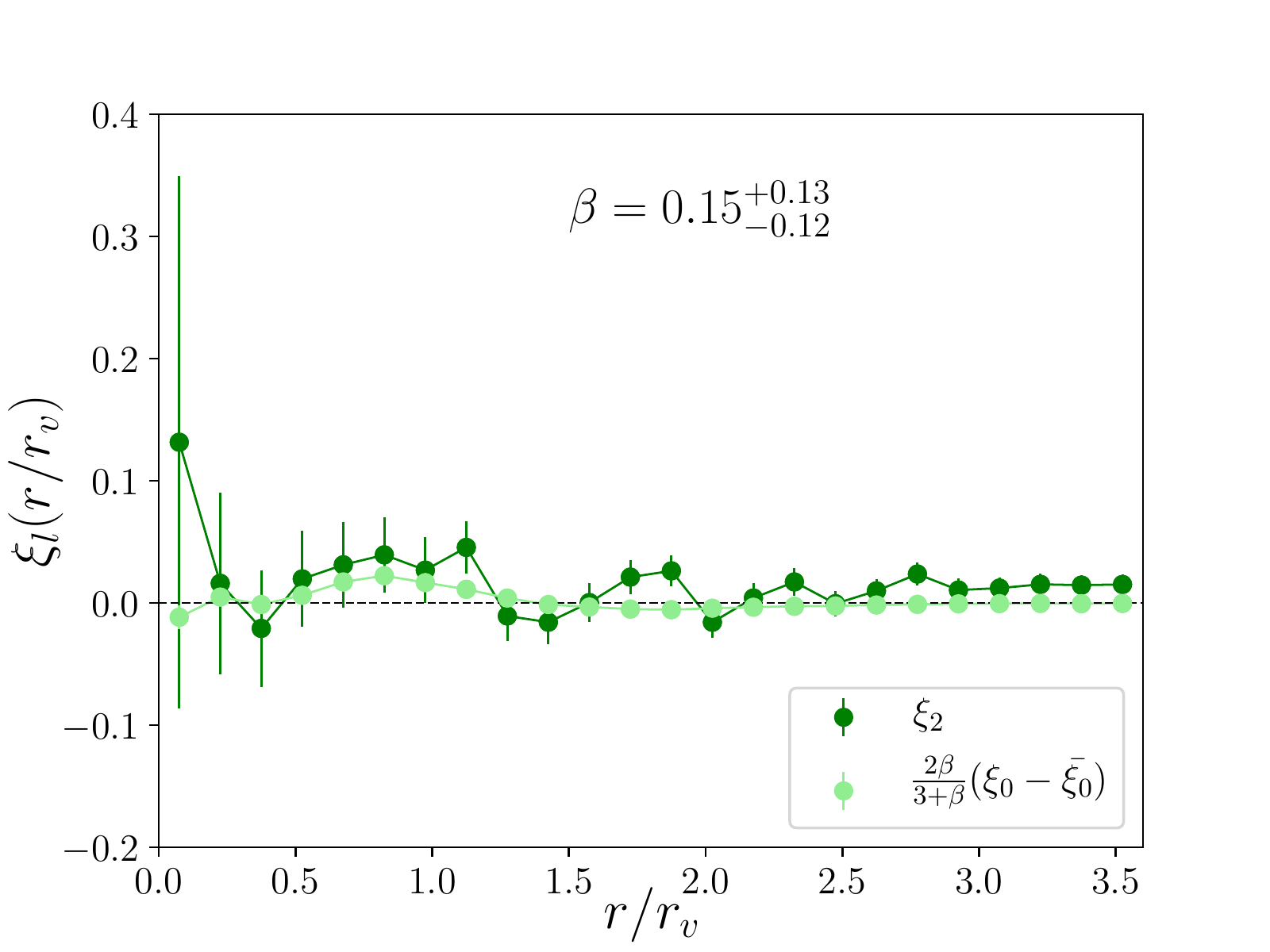} 
        \caption{QSO}
        \label{fig:QSObeta}
    \end{subfigure}
\caption{Quadrupole (dark green) of the void-galaxy cross-correlation function for LRGs, subpanel (a), and QSOs, subpanel (b). Error bars are derived from the diagonal elements of the covariance matrices. Also plotted is the best fitting model quadrupole (light green).}
  \end{figure}

Figure \ref{fig:LRGbeta} shows the quadrupole for voids in the LRG catalogue, after the MVA cut has been applied. 
Also shown in this figure is the quantity $\frac{2\beta}{3+\beta}(\xi_0-\bar{\xi}_0)$ for the best fitting value of $\beta= 0.58 ^{+0.33}_{-0.28}$. In our model, this quantity should be equal to the quadrupole. 
The quadrupole is less pronounced than the monopole, though still significantly non-zero around $r/r_v \approx 1$.

Assuming the value for the bias of the LRGs quoted in \cite{Zhai2017}, and our fiducial value of $\sigma_8$, we therefore find that $f\sigma_8 = 0.76^{+0.43}_{-0.37}$. This agrees, to within one sigma, with the value of $f\sigma_8$ in our fiducial cosmology at the mean redshift of the void sample of $\bar{z} = 0.703$, which is $f\sigma_8(z=0.703) = 0.47$.  

Figure \ref{fig:QSObeta} shows the measured quadrupole for the observed void-galaxy cross-correlation function for voids in the QSO catalogue. Also shown is the quantity $\frac{2\beta}{3+\beta}(\xi_0-\bar{\xi}_0)$ for the best fitting value of $\beta= 0.15 ^{+0.13}_{-0.12}$. The quadrupole is much less pronounced than the monopole. It is difficult to see by eye whether or not there is a significant quadrupole. Indeed, the quadrupole is quite flat and is consistent with zero at most scales. 

Assuming the value for the bias of the QSOs quoted in \cite{Laurent2017}, and our fiducial value of $\sigma_8$ at the mean redshift of the void sample of $\bar{z} = 1.53$, we find that $f\sigma_8 = 0.14 ^{+0.12}_{-0.11}$. Although a growth rate of zero is excluded, our recovered value does not agree, to within one sigma, with the value of $f\sigma_8$ in our fiducial cosmology, which is $f\sigma_8(z=1.53) = 0.37$, though it is less than $2\sigma$ away. If the QSO void catalogue is indeed heavily contaminated by spurious voids (as suggested by the MVA analysis) then this would explain why we appear to under estimate the growth rate (since spurious voids bias low the quadrupole average value).

Table \ref{table:stats2} summarises our measurements of the growth rate of structure around voids in the LRG and QSO samples.


\section{Discussion}
\label{sec:discussion}

In this paper we have elucidated 
our search for voids in DR14 of eBOSS. We identified voids in the LRG and QSO galaxy redshift catalogues. We applied a multivariate analysis to these void catalogues in an attempt to expunge Poissonian voids contaminating the sample. By applying a model for the monopole and quadrupole of the void-galaxy cross-correlation function, based on linear RSD theory, we were able to estimate the growth rate $\beta$ around voids in these catalogues. The cross-correlation between voids identified in the LRG catalogue shows strong evidence for a positive value of $\beta$. Our estimate of $\beta = 0.58 ^{+0.33} _{-0.28}$ from this sample is consistent with other values of the growth rate presented in the literature. 

However, we have been unable to show that the cross-correlation between voids in the QSO catalogue and the QSO tracers can be used to make a significant measurement of the growth rate of structure. Our measurement of $\beta = 0.15 ^{+0.13} _{-0.12}$ is much lower than expected from our fiducial cosmology and other values of the growth rate in the literature. Our application of MVA techniques provides a possible explanation for this. By comparing voids in mock QSO catalogues with voids found in random catalogues we showed that the QSO void sample is highly contaminated with spurious Poisson voids. This is due to the sparsity of the QSO redshift catalogue. Our current methodology is not capable of separating Poissonian voids from true voids in the QSO catalogue. These Poisson voids suppress the measured quadrupole and thus attenuate the strength of the RSD signal. The value of the work presented here therefore lies in the fact that we are able to make a quantitative statement about the reliability and the usability of our void catalogues.

It has been suggested by reference \cite{NadathurPercival2017} that the RSD model for the multipoles of the void-galaxy cross-correlation function that we have applied here could be extended. The extended model they present contains terms proportional to the real space matter density profile of the void, a quantity which is \textit{a priori} unknown. Therefore, they argue that the monopole to quadrupole ratio estimator of the growth rate is only approximate, and depends on the validity of the assumptions made in deriving the model. However, differences between the model used here and their model are most apparent on scales $r<r_v$, the scales where the uncertainty of the measurement of the multipoles of the cross-correlation is largest. Given these large errors any systematic bias introduced by the RSD model is not measurable. Other differences between our methodology and that presented in Ref. \cite{NadathurPercival2017}, such as the void centre definition and the stacking method, prevent us from fully assessing the importance of including higher order terms. Therefore the decision not to apply the extended model here is a valid one. That being said, in future work where we will be dealing with smaller statistical errors, the difference between the model applied here and that in \cite{NadathurPercival2017} can be thouroughly investigated.
 
eBOSS has now finished collecting data. The final galaxy clustering catalogues for the LRG, ELG, and QSO samples have been built. In the near future we shall repeat our analysis on this larger final sample. The final eBOSS void catalogues will have a higher number of void-galaxy pairs and will thus reduce the statistical errors on measurement of the void-galaxy cross-correlation. With more precise measurement of the growth rate of structure around voids at high redshift we will be able to investigate the redshift dependence of the growth rate. We will be able to provide the most reliable ever measurement of the growth rate of structure with voids at these redshifts. Furthermore, completion of the ELG catalogue will allow us to search for voids with those tracers too. We will also be able to cross correlate voids identified in the ELG catalogue with tracers in the LRG catalogue and {\it vice versa}. A more complete sample will also enable us to measure these sorts of cross-correlations anisotropically, and thus provide further estimates of the growth rate of structure.

\acknowledgments

We thank the support of the OCEVU Labex
(Grant No ANR-11-LABX-0060) and the A*MIDEX
project (Grant No ANR-11-IDEX-0001-02) funded by
the Investissements d\textquotesingle Avenir French government program managed by the ANR. We also acknowledge support from the ANR eBOSS project (ANR-16-CE31-0021) of the French National Research Agency.

AP is supported by NASA grant 15-WFIRST15-000-8 to the WFIRST Science Investigation Team "Cosmology with the High Latitude Survey".

G.R. acknowledges support from the National Research Foundation of Korea (NRF)
through Grant No. 2017077508 funded by the Korean Ministry of Education, Science
and Technology (MoEST), and from the faculty research fund of Sejong University in
2018.

Funding for the Sloan Digital Sky Survey IV has been provided by the Alfred P. Sloan Foundation, the U.S. Department of Energy Office of Science, and the Participating Institutions. SDSS acknowledges support and resources from the Center for High-Performance Computing at the University of Utah. The SDSS web site is www.sdss.org.

SDSS-IV acknowledges support and resources from the Center for High-Performance Computing at the University of Utah.  

SDSS is managed by the Astrophysical Research Consortium for the Participating Institutions of the SDSS Collaboration including the Brazilian Participation Group, the Carnegie Institution for Science, Carnegie Mellon University, the Chilean Participation Group, the French Participation Group, Harvard-Smithsonian Center for Astrophysics, Instituto de Astrof\'isica de Canarias, The Johns Hopkins University, Kavli Institute for the Physics and Mathematics of the Universe (IPMU) / University of Tokyo, the Korean Participation Group, Lawrence Berkeley National Laboratory, Leibniz Institut f\"ur Astrophysik Potsdam (AIP), Max-Planck-Institut f\"ur Astronomie (MPIA Heidelberg), Max-Planck-Institut f\"ur Astrophysik (MPA Garching), Max-Planck-Institut f\"ur Extraterrestrische Physik (MPE), National Astronomical Observatories of China, New Mexico State University, New York University, University of Notre Dame, Observat\'orio Nacional / MCTI, The Ohio State University, Pennsylvania State University, Shanghai Astronomical Observatory, United Kingdom Participation Group, Universidad Nacional Aut\'onoma de M\'exico, University of Arizona, University of Colorado Boulder, University of Oxford, University of Portsmouth, University of Utah, University of Virginia, University of Washington, University of Wisconsin, Vanderbilt University, and Yale University.




\bibliographystyle{JHEP}
\bibliography{eBOSSvoidRSD}







\end{document}